\documentclass[sigconf, nonacm]{acmart}

\newcommand{\revision}[1]{{\leavevmode\color{black}{#1}}}
\usepackage{xparse}
\usepackage{balance}
\usepackage{mathtools}
\usepackage{tikz}
\usepackage{tabularx}
\usepackage{booktabs} 
\usepackage{ragged2e}
\usepackage{subcaption}
\usepackage{xcolor,pifont, colortbl}
\usepackage{graphics}
 \usepackage{enumitem}
\usepackage[export]{adjustbox}
\usepackage[a-2b]{pdfx}
\usepackage{graphicx}

\usepackage{tikz}
\newcommand*\circled[1]{\tikz[baseline=(char.base)]{
            \node[shape=circle,fill,inner sep=1.3pt] (char) {\textcolor{white}{#1}};}}

\definecolor{Gray}{gray}{0.85}
\definecolor{LightCyan}{rgb}{0.88,1,1}
\definecolor{lightblue}{RGB}{170,200,255}
\definecolor{lightgreen}{RGB}{130,255,200}
\newcolumntype{a}{>{\columncolor{Gray}}l}
\newcolumntype{b}{>{\columncolor{white}}l}
\newcommand*\colourcheck[1]{
  \expandafter\newcommand\csname #1check\endcsname{\textcolor{#1}{\ding{52}}}%
}
\colourcheck{blue}
\colourcheck{green}
\colourcheck{red}
\newcommand*\colourmark[1]{
  \expandafter\newcommand\csname #1mark\endcsname{\textcolor{#1}{\ding{55}}}%
}
\colourmark{red}

\NewDocumentCommand\DownArrow{O{2.0ex} O{black}}{%
   \mathrel{\tikz[baseline] \draw [<-, line width=0.5pt, #2] (0,0) -- ++(0,#1);}
}

\newcommand\vldbdoi{10.14778/3598581.3598585}
\newcommand\vldbpages{2103 - 2116}
\newcommand\vldbvolume{16}
\newcommand\vldbissue{9}
\newcommand\vldbyear{2023}
\newcommand\vldbauthors{\authors}
\newcommand\vldbtitle{\shorttitle} 
\newcommand\vldbavailabilityurl{URL_TO_YOUR_ARTIFACTS}
\newcommand\vldbpagestyle{plain} 

\begin{document}
\title{WiscSort: External Sorting For Byte-Addressable Storage}


\author{Vinay Banakar}
\affiliation{%
  \institution{University of Wisconsin Madison}
}
\email{vin@cs.wisc.edu}

\author{Kan Wu}
\affiliation{%
  \institution{Google}
}
\email{kanwu@google.com}

\author{Yuvraj Patel}
\affiliation{%
  \institution{University of Edinburgh}
}
\email{yuvraj.patel@ed.ac.uk}

\author{Kimberly Keeton}
\affiliation{%
  \institution{Google}
}
\email{kimkeeton@google.com}

\author{Andrea C. Arpaci-Dusseau}
\affiliation{%
  \institution{University of Wisconsin Madison}
}
\email{dusseau@cs.wisc.edu}

\author{Remzi H. Arpaci-Dusseau}
\affiliation{%
  \institution{University of Wisconsin Madison}
}
\email{remzi@cs.wisc.edu}

\begin{abstract}
We present WiscSort, a new approach to high-performance concurrent sorting for existing and future byte-addressable storage (BAS) devices. 
WiscSort carefully reduces writes, exploits random reads by splitting keys and values during sorting, and performs interference-aware scheduling with thread pool sizing to avoid I/O bandwidth degradation. 
We introduce the $BRAID$ model which encompasses the unique characteristics of BAS devices. 
Many state-of-the-art sorting systems do not comply with the BRAID model and deliver sub-optimal performance, whereas WiscSort demonstrates the effectiveness of complying with BRAID.
We show that WiscSort is 2-7x faster than competing approaches on a standard sort benchmark.
We evaluate the effectiveness of key-value separation on different key-value sizes and compare our concurrency optimizations with various other concurrency models.
Finally, we emulate generic BAS devices and show how our techniques perform well with various combinations of hardware properties.
\end{abstract}

\maketitle

\pagestyle{\vldbpagestyle}
\begingroup\small\noindent\raggedright\textbf{PVLDB Reference Format:}\\
\vldbauthors. \vldbtitle. PVLDB, \vldbvolume(\vldbissue): \vldbpages, \vldbyear.\\
\href{https://doi.org/\vldbdoi}{doi:\vldbdoi}
\endgroup
\begingroup
\renewcommand\thefootnote{}\footnote{\noindent
This work is licensed under the Creative Commons BY-NC-ND 4.0 International License. Visit \url{https://creativecommons.org/licenses/by-nc-nd/4.0/} to view a copy of this license. For any use beyond those covered by this license, obtain permission by emailing \href{mailto:info@vldb.org}{info@vldb.org}. Copyright is held by the owner/author(s). Publication rights licensed to the VLDB Endowment. \\
\raggedright Proceedings of the VLDB Endowment, Vol. \vldbvolume, No. \vldbissue\ %
ISSN 2150-8097. \\
\href{https://doi.org/\vldbdoi}{doi:\vldbdoi} \\
}\addtocounter{footnote}{-1}\endgroup

\ifdefempty{\vldbavailabilityurl}{}{
\vspace{.3cm}
\begingroup\small\noindent\raggedright\textbf{PVLDB Artifact Availability:}\\
The source code, data, and/or other artifacts have been made available at \url{https://gitfront.io/r/vinaybanakar/tHq3i8MVUk9q/WiscSort}.
\endgroup
}
\section{Introduction}
External sorting is a critical component of many modern data intensive applications (web indexing \cite{wsc}, key-value stores \cite{leveldb, rocksdb}, data analytics \cite{wsc}, and  relational databases \cite{postgress-sort, duckdb, sqlite-sort}), making use of external storage to sort data that does not fit in DRAM. For example, relational databases (such as MySQL and PostgresSQL) use external sorting for \texttt{ORDER BY} queries on non-indexed keys or to handle \texttt{TOP-K} queries whose input exceeds the available memory.

Traditionally HDDs or SSDs have been used for external storage; however, Byte-Addressable Storage (BAS) is emerging as an appealing expansion memory or fast storage layer for data-intensive applications \cite{samsung-cxl-expander, kioxia-cxl, tpp, first-gen-cxl}. This broad class of storage devices includes the recently announced Samsung CXL Memory-Semantic SSD \cite{samsung-memory-ssd} and Intel's Optane DC Persistent Memory (PMEM) \cite{intel-optane-pmm}. BAS devices provide a larger capacity than DRAM and are considerably faster than traditional SSDs and HDDs, making them suitable for performance-critical services. 

For applications to maximize performance, the unconventional characteristics of BAS devices must be carefully considered. We introduce a generic device model called $BRAID$ that depicts the typical performance characteristics of byte-addressable storage devices. The BRAID model has five crucial properties: \textbf{B}yte addressability, \textbf{R}andom read performance, \textbf{A}symmetric read-write cost, read-write \textbf{I}nterference, and \textbf{D}evice constrained concurrency. Combinations of two or more of these characteristics can be seen in a BRAID device; for example, PMEM exhibits all five BRAID properties.


This paper presents \textit{WiscSort}, a BRAID-conscious high-performance external sorting algorithm derived from the popular external merge sort.
WiscSort postpones the movement of a (key, value) pair's value until the pair's final sorted position is known (typically, the final merge phase of external merge sort); only the keys are moved between DRAM and the storage device during earlier phases.
Except for the final phase, WiscSort maintains keys and pointers in DRAM, whereas previous approaches bundled keys and values. 
Pointers point to the respective values in the original file on the device.
This simple late materialization avoids writes of values during early sorting phases; for popular workloads with keys smaller than values \cite{benchmark-rocksdb-fb}, the savings are significant.
Furthermore, because WiscSort only keeps pointers in DRAM, it can generate larger sorted runs during the run-generation phase, reducing the number of merge phases or avoiding the merge phase completely.

BRAID devices can also have peculiar concurrency characteristics. For example, writes do not scale well, and the read bandwidth degrades (up to 2x \cite{emperical-guide-optane}) when there are overlapping write requests. 
To alleviate these read-write interference degradation and concurrency constraints, we introduce a \emph{Thread-Pool Controller} and an \emph{Interference-Aware Scheduler}.
WiscSort utilizes the thread-pool controller to determine the appropriate concurrency pool size (for read/write) based on the access type, and the interference-aware scheduler schedules reads and writes to the device in a non-overlapping fashion to maximize performance.


We compare the performance of WiscSort with classic external merge sort implementations on both microbenchmarks and standard application level benchmarks.
We show that WiscSort performs $2x$--$3x$ better than concurrent external merge sort, $5x$ better than state-of-the-art in-place sample sort, and $7x$ better than recent PM based sorting system (PMSort) on sortbenchmark \cite{sortbenchmark} workloads.
The interference-aware scheduler and the thread-pool controller reduce total time by at least $50$\% compared to approaches that overlap reads and writes. 
Moreover, a system that just separates keys and values but is not aware of these concurrency properties is $\sim$15\% slower than one that is interference and concurrency aware.
We show that special cases of WiscSort can do better than external merge sort even when the value size is smaller than the key size. 
As the value size increases, the performance gap between merge sort and WiscSort grows. 
We also demonstrate that using random reads to reduce unnecessary data loading is better than sequentially reading all data.
Finally, we project the performance of WiscSort on emulated BRAID devices and discuss the benefits of our techniques.

\section{Background and Motivation}

This section provides background on external sorting and new byte-addressable storage. We explain why existing sorting approaches for DRAM/HDD/SSD are insufficient for BRAID devices.


\subsection{Traditional External Sorting}
\label{externalsort}
External sorting refers to a class of algorithms capable of sorting large amounts of data that do not fit the available memory.
Among various external sorting algorithms, external merge sort is a popular approach~\cite{Knuth1973, sort-io-complexity, tritonsort}. 
External merge sort sorts data in two phases: run generation and merge. In the run generation phase, data chunks are read into DRAM, sorted, and written to a temporary run file.
In the merge phases, the sorted run files are then combined; large amounts of data or small DRAM sizes may necessitate multiple merge phases since a record from each run file might not fit in available memory. Traditional external merge sort only performs sequential I/O as it is designed for SSD/HDDs.

\subsection{Byte-Addressable Storage (BAS)}

Blocks have been the traditional unit of I/O for mass storage systems, and data management systems (including the operating system) have been heavily optimized for block-based device characteristics. However, the introduction of byte-addressable storage devices requires software to be redesigned for their new properties. 
Two simple ways of using these new devices are to treat them as faster disks or as slower DRAM; we explore the pitfalls of doing so.


In the past, many persistent memory technologies have been proposed, such as Phase Change Memory \cite{PCM}, STT-RAM \cite{sttRAM}, and ReRAM \cite{reRAM}, but the release of Intel Optane DC PMEM in 2019 ushered in an era of new programming models, interfaces \cite{Mnemosyne, ddp}, file systems \cite{splitfs, nova}, caching systems \cite{nyxcache}, and database indexes \cite{lbtree, utree}.  PMEM is a byte-addressable storage device that can be accessed via hardware load/stores through the on-chip memory controller and is placed on the DIMM slots; PMEM can be configured in Memory (direct-mapped cache) or AppDirect mode (BAS).


According to the Intel 2022 Q2 earning release \cite{optanewinddown}, Intel is discontinuing its Optane Memory business. However, the problems PMEM originally aimed to address, such as DRAM capacity scaling and slow persistence, remain unsolved. 
Optane being the only persistent memory available in the market currently puts many software solutions that adopted Optane at risk \cite{sap-nvm, oracle-nvm, vmware-nvm}, but this is expected to be alleviated by the introduction of new byte-addressable storage devices.

Several byte-addressable storage devices, such as the Samsung CXL Memory Semantic SSD~\cite{samsung-memory-ssd}, Kioxia CXL 3D flash memory~\cite{kioxia-cxl}, and Everspin STT-RAM \cite{everspin-stt}, are set to be released soon, due to the growing industry's transition towards the CXL standard~\cite{cxl}. CXL, a cache-coherent interconnect that runs on top of PCIe, enables memory expanders such as the Samsung 512GB module \cite{samsung-cxl-expander} to increase system memory capacity without consuming DIMM slots. Prototypes from Samsung~\cite{enabling-cxl-db}, KAIST \cite{directCXL, first-gen-cxl} and Meta \cite{tpp} have demonstrated the potential of CXL technology. CXL and JDEC (DDR)~\cite{jdec} memory standards have agreed to collaborate on new persistent memory research, strengthening the availability of future BAS devices~\cite{cxl-jdec}. With BAS expected to return over CXL, it is believed that properties such as byte addressability, improved random-read performance, and read-write asymmetry will still remain. Preliminary results show similar latencies to that of accessing remote socket memory, with access latency of 230 ns and 32 GB/s bandwidth on a PCIe Gen 5.0 \cite{samsung-cxl-expander}.

\subsection{The BRAID model}
\label{key-device-properties}
We develop a device model for byte-addressable storage that specifies the important properties that distinguish it from other storage media. Being cognizant of these unconventional properties is crucial for maximizing performance. The BRAID model constitutes a device with the following five properties:
\begin{enumerate}[leftmargin=0.5cm]
 \item \textbf{Byte Addressability (\textit{B}).} As BAS is byte addressable, it allows access to small data regions without the amplification of page granularity requests. This property helps reduce unnecessary data movement over the memory bus and thus does not waste bandwidth unnecessarily.

 \item \textbf{Higher Random-Read performance (\textit{R}).} Random-read performance on BAS is on par with sequential read performance for larger accesses \cite{emperical-guide-optane}. Concurrent random reads are only $~18$\% slower than concurrent sequential reads for 256B accesses on PMEM.

 \item \textbf{Asymmetric Read-Write Cost (\textit{A}).} There is a vast difference between read and write performance on BAS. Read performance on PMEM is up to $4x$ better than write performance \cite{emperical-guide-optane}. 
 \revision{Previous devices like HDDs, DRAMs, and SSDs exhibit similar asymmetry but at different degrees.}
\item \textbf{Read-Write Interference (\textit{I}).} Read performance degrades when concurrent writes are issued on BAS. This degradation increases with an increased number of concurrent writes. However, the opposite effect is minimal, i.e., there is little to no degradation in write performance when multiple reads are performed concurrently \cite{nyxcache}.

\item \textbf{Device-Constrained Concurrency (\textit{D}).} BAS exhibits specific concurrency constraints; for example, writes do not scale well, but reads do until the number of read threads matches the total physical cores. Performing writes with the maximum number of threads can be $\sim$2x slower than peak write performance \cite{olap-nvm}.

\end{enumerate}
\revision{
We expect these properties to remain prevalent in BAS devices for the foreseeable future, but we also examine devices without some of these common characteristics. 
BRAID serves as a comprehensive model encompassing various devices with different characteristics, although not all devices may possess all of these traits. 
For example, future devices might have different constraints of concurrency (\textit{D}) or might not exhibit interference effects at all (\textit{I} = 0), yet they can still be represented by the BRAID model; of course, the methods required to fully utilize the device (be "BRAID compliant") may differ.
In the rest of the paper, we sometimes refer to a BRAID device simply as BRAID for brevity.}

\subsection{The Question: How to Sort on BRAID?}
Existing sorting algorithms, unfortunately, do not readily translate into efficient BRAID solutions. In the following, we look at two types of sorting on BRAID: 1) in-place sorting with the BRAID device treated as a slower DRAM, and 2) external sorting with BRAID treated as a faster HDD/SSD. We demonstrate why using these two existing approaches to sort on a BRAID device is inefficient.

\vspace{-1.5mm}
\subsubsection{\textbf{BRAID As A Slower DRAM}}
\label{samplesort}
A BRAID device is an order of magnitude slower than DRAM, making direct in-place sorting on PM inefficient. In-place sorting algorithms, such as sample sort, move records around based on pairwise record comparisons. These algorithms produce $log_2{N}$ times the record movement traffic normalized to the dataset size. When sorting in-place directly on BRAID without using DRAM, all of the traffic translates into slow BRAID accesses.
In contrast, if we use external sorting algorithms, a significant portion of the traffic will be served by fast DRAM, reducing total sorting time significantly. External merge sort on BRAID, for example, produces $(1+M)$ times the dataset size of BRAID traffic (M is the number of merge phases, $M=1$ in dominant cases).

\begin{figure}[!t]
    \begin{minipage}{0.47\textwidth}
    \centering
    \includegraphics[width=0.88\textwidth]{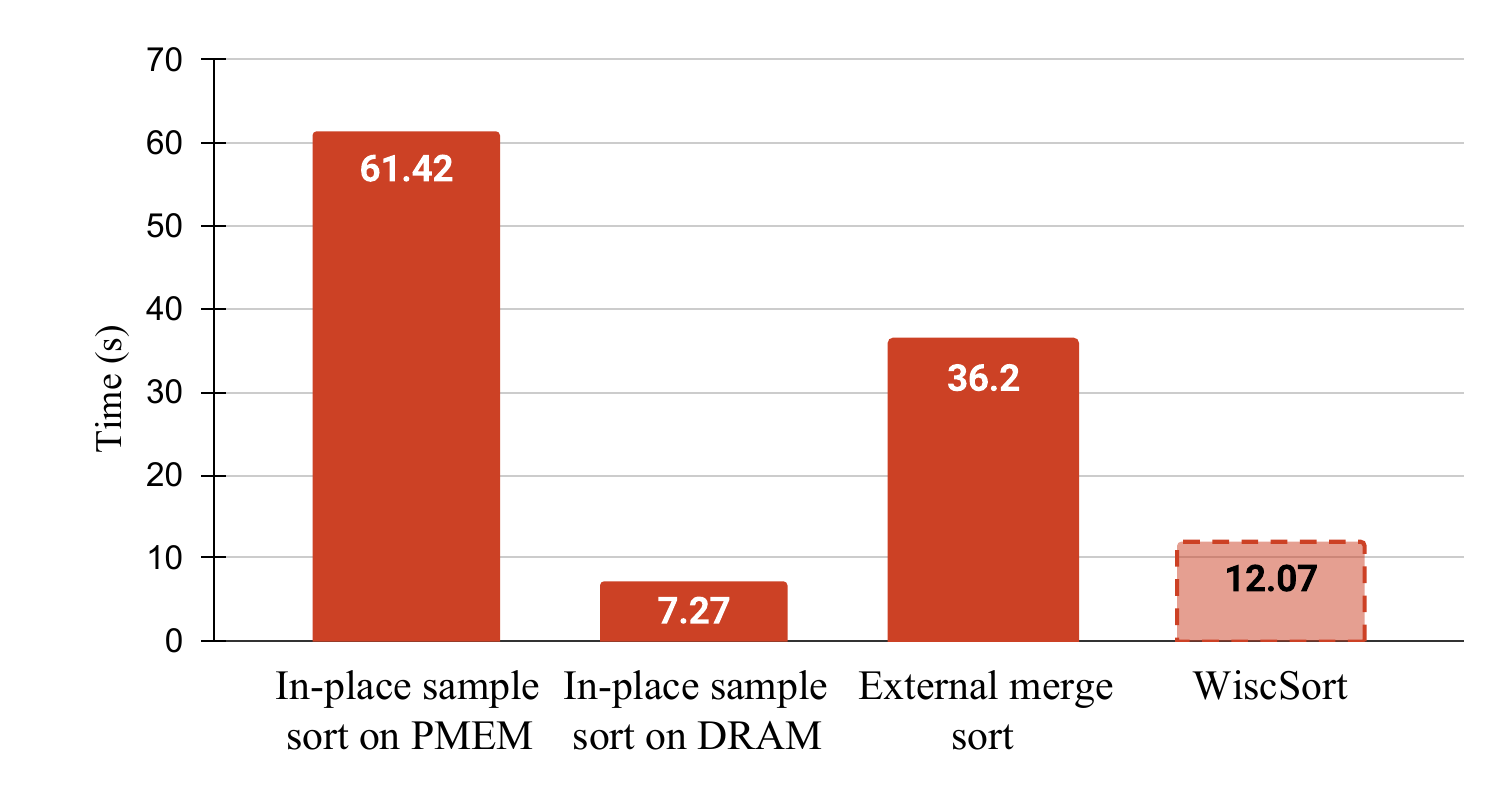}
    \vspace{-3mm}
    \caption{Problems of different sorting approaches on PMEM. \textmd {We plot sorting time of 1) In-place sample sort, 2) Traditional external merge sort, and 3) our WiscSort on PMEM for a 20GB workload containing 200M records with 10B keys and 90B values.}}
    \label{fig:WiscSort - motivation}
    \end{minipage}
    \vspace*{-4.45mm}
\end{figure}

A state of the art in-place concurrent sample sort \cite{IPS4o} fails to consider the concurrency constraints or the interference properties of the BRAID model, as it was designed for DRAM.
As shown in Figure \ref{fig:WiscSort - motivation}, external merge sort performs $\sim$2x faster than in-place sample sort, as it requires much less device traffic and respects the device concurrency properties in comparison. Moreover, in-place sorting on DRAM is $\sim$10x faster than in-place sorting on PMEM. Hence, we conclude that in-place sorting on PM is inefficient.

\vspace{-1.5mm}
\subsubsection{\textbf{BRAID As A Faster Disk.}}
Previous external sort implementations designed for HDD/SSD are also insufficient on BRAID, particularly in terms of value movement during sorting.
Assume the dataset consists of key-value pairs. External sorting implementations typically move values along with keys, even though sorting only involves key comparisons. For example, in external merge sort, both keys and values are read into DRAM and written to temporary BRAID files during run generation; similarly, both keys and values are read and written during merge.

Moving values with keys is effective because it leverages the sequential operation of HDD/SSDs. Random reads on HDD/SSD are $\sim$1.5x slower than sequential reads (unlike $(R)$). Furthermore, the 4KB access granularity of HDD/SSDs is much larger than popular sorting workload value sizes (e.g., 100B in GraySort) (unlike $(B)$). If we do not move values with keys on HDD/SSDs, we save sequential read/write of values but introduce slow random reads to fetch values into their sorted positions, and each random read yields large amplification (40x = 4KB/100B in case of GraySort). Because of these HDD/SSD characteristics, moving values with keys is advantageous in external sorting (unlike $(A)$). Traditional concurrent external sorts \cite{parallel-external-sort} only consider merge based parallelism or partition based parallelism and are ignorant of the device based parallelism required on modern storage (unlike $(I, D)$) for maximum performance.

However, unlike HDD/SSD, BRAID devices have fundamentally different performance characteristics, necessitating this tradeoff to be reconsidered. 
BRAID has a limited write bandwidth while providing excellent random-read performance (e.g., a single PMEM DIMM has 2.5GB/s sequential write vs 7GB/s random read bandwidth).
Because BRAID supports fine-grained access (256B for PMEM), small random reads required by sorting workloads become significantly more efficient on BAS when compared to HDD/SSD.
As we will demonstrate, due to these unique characteristics, existing data movement schemes that do not comply with the BRAID model leave the true potential of BAS largely underutilized.

\vspace{-2mm}
\subsubsection{\textbf{Separating Key from Values.}}
Separating the key and value to improve sorting is a classic idea. A 1963 CACM paper \cite{cacm63} proposed separating keys from values to perform just the key-pointer sort; however, due to the slow random reads on hard drives, they convert all random reads to sequential reads for gathering the values, thus performing more sorts than required. Moreover, they fail to address the I/O amplification of using block accesses when processing small keys. We examine this six-decade-old approach for modern hardware, which is byte addressable and has random bandwidth reaching near-sequential bandwidths.

PMSort \cite{pmsort} performs key-value separation for PMEM to reduce write traffic, focusing on the single-threaded case.
However, it does not fully exploit BRAID properties and makes some choices that do not scale. 1) PMSort does not fully take advantage of the random-read bandwidth, as it loads both keys and values to the memory during the RUN phase. 2) They conclude QuickSort is the best approach for sorting on PMEM, but as we will show (see Figure \ref{fig:WiscSort - motivation}), this does not scale. 3) PMSort avoids performing random reads (like \cite{cacm63}) and claims bandwidth not to be the bottleneck, which is not true at scale. 4) PMSort focuses on wear-leveling, thus not fully utilizing all the properties of a BRAID device. 



\begin{table}
\tiny
\caption{Sorting system's compliance with the BRAID model.}
\vspace*{-2.5mm} 
\centering
\resizebox{\columnwidth}{!}{
\begin{tabular}{accccc}
\hline
\rowcolor{lightblue}
System & B & R & A & I & D\\ \hline
External merge sort (naive) & \redmark & \redmark & \redmark & \redmark & \redmark \\ \hline
In-place sample sort \cite{IPS4o} & \greencheck & \greencheck & \redmark  & \redmark & \redmark \\ \hline
External merge sort  & \redmark & \redmark & \redmark & \greencheck & \greencheck \\ \hline
Modified-key sort \cite{cacm63} & \redmark & \redmark & \greencheck & \redmark & \redmark \\ \hline
PMSort \cite{pmsort} & \greencheck & \redmark & \greencheck & \redmark & \redmark \\ \hline
WiscSort     & \greencheck & \greencheck & \greencheck & \greencheck & \greencheck \\ \hline
\end{tabular}
} 
\label{tab:property}
\vspace{-8pt} 
\end{table}

Table \ref{tab:property} summarizes how different sorting systems adhere to the BRAID model. Traditional external merge sort is not device concurrency constraint aware, but we add a thread pool controller (Sec \ref{tpool-control}) and interference-aware scheduling (Sec \ref{IAS}) for it to be a competitive comparison against WiscSort.
Modified-Key Sort makes a conscious decision to avoid random reads due to the exorbitant cost on older devices. 
PMSort does not completely take advantage of random-read performance and does not attempt to be device concurrency or interference aware. 
WiscSort, in contrast, is a practical real-world sorting system that takes advantage of all the properties specified in the BRAID model for maximum performance.

\subsection{Our target workload}
\label{assumptions}
Based on the observations, we aim to design a system that efficiently sorts large volumes of input data using byte-addressable storage. As row-oriented binary data formats become relevant again \cite{databricksDataLakehouse, apacheAvro, googleDataLake}, we rely on the format specified by sortbenchmark \cite{sortbenchmark}, a well-known sorting benchmark designed to stress test the I/O subsystem. Specifically, the workload has uniformly random keys, it is read from and written to files on BRAID, and the size of keys and values are fixed for a dataset. 
\revision{For variable length values, we rely on Key-Length-Value encoding \cite{klv}, where a fixed size key is followed by the length of the value and the value itself. This simple and widely used record format (SQLite \cite{sqlite-format}, PostgreSQL \cite{pg-format}, etc.) allows us to make no assumptions about the index data structures.}
Additionally, we assume the BRAID capacity is large enough to fit the dataset and intermediate run files.

\section{WiscSort}
In the previous section, we discussed how byte-addressable storage has different properties compared to slower counterparts such as HDDs or SSDs. Using these properties, we introduce WiscSort, a new algorithm that performs external merge sort on BRAID.
WiscSort is a single-machine sorting algorithm that exploits the properties of the BRAID model (\ref{key-device-properties}) to efficiently utilize the higher bandwidth offered. Although the properties of the $BRAID$ model are derived from PMEM, we expect most of them to also be present in future storage devices. 


\subsection{Design goals}
Based on the BRAID model, we derive five high-level goals to achieve high throughput:
\begin{itemize}[leftmargin=0.6cm]
 \item \textbf{Reduce I/O amplification.} Small accesses to storage devices need no longer be amplified due to the byte granularity offered. Since BRAID bandwidth can be saturated even when making small accesses, reducing total I/O traffic by making byte-level accesses should be preferred.

 \item \textbf{Avoid redundant reads.} Conventionally, systems focus on maximizing bandwidth using sequential accesses; however, this approach is no longer required due to BRAID's high random-read bandwidth.

 \item \textbf{Trade more reads for fewer writes.} Current BRAID devices have asymmetric read-write costs; thus, trading more reads for fewer writes to maximize performance is preferred.

\item \textbf{Manage access concurrency.} Spawning too few threads or too many threads can hurt performance on BRAID. Hence, we must appropriately size the thread pool based on the access type and the device to maximize I/O bandwidth. 

 \item \textbf{Avoid read-write interference.} Overlapping read and write workloads on BRAID can lead to reduced read bandwidth. Thus, we must ensure that reads and writes are not overlapped.
\end{itemize}

\subsection{Overview}

\begin{figure*}
\centering
\begin{subfigure}{.31\textwidth}
  \captionsetup{justification=centering}
  \centering
    \includegraphics[width=0.90\textwidth]{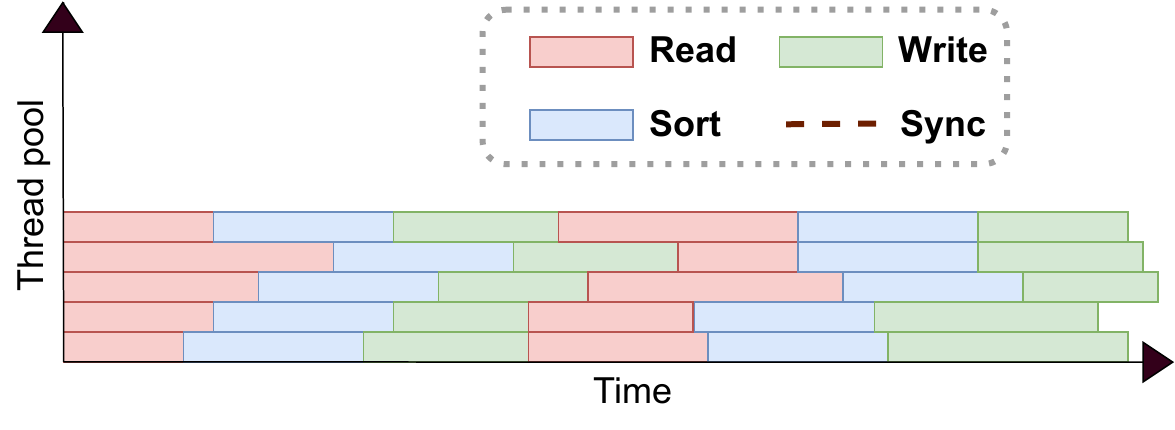}
    \caption{\footnotesize No Synchronization}
    \label{fig:nosync}
\end{subfigure}%
\begin{subfigure}{.30\textwidth}
  \captionsetup{justification=centering}
  \centering
    \includegraphics[width=0.90\textwidth]{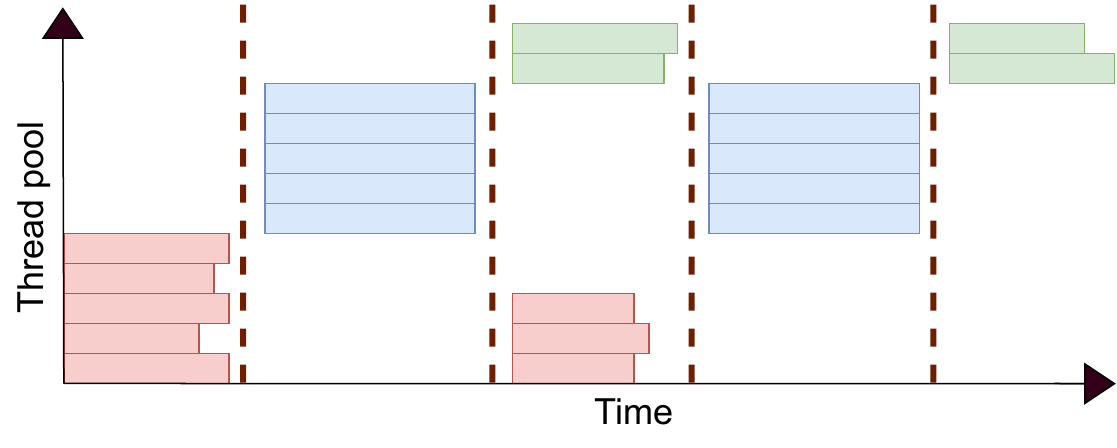}
    \caption{ \footnotesize R/W Thread Pools}
    \label{fig:overlapIO}
\end{subfigure}%
\begin{subfigure}{.35\textwidth}
  \captionsetup{justification=centering}
  \centering
    \includegraphics[width=0.90\textwidth]{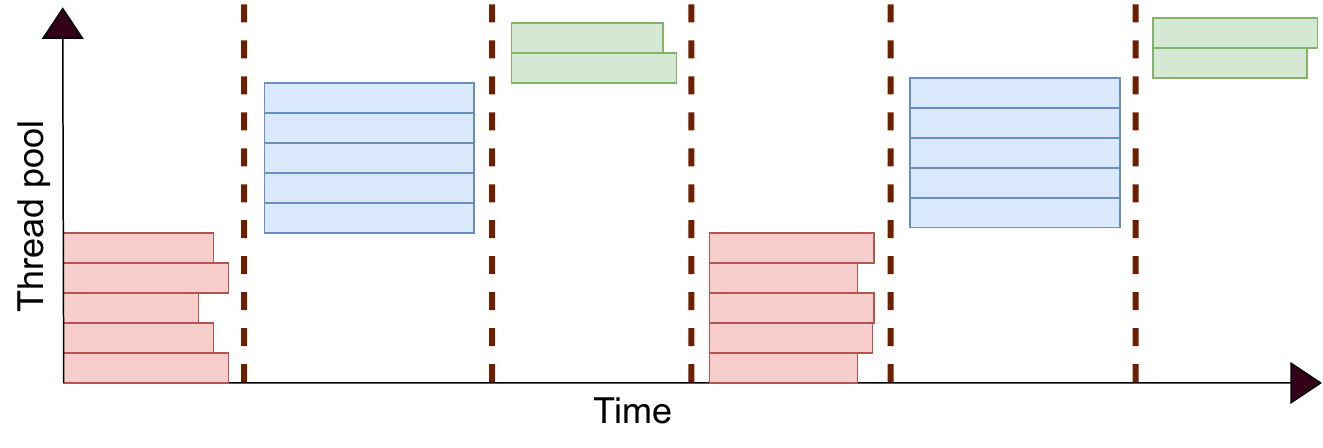}
    \caption{\footnotesize R/W Thread Pools w/ Non-Overlapping Requests}
    \label{fig:nooverlapIO}
\end{subfigure}
\vspace{-3mm}
\caption{
    \revision{
        Three different concurrency mechanisms. \textmd{
            To maximize BRAID device bandwidth, WiscSort prefers \ref{fig:nooverlapIO} model (interference aware, thread-pool controller) over \ref{fig:overlapIO} (thread-pool controller) and \ref{fig:nosync} (interference unaware).
            In all three models, the sort pools are the same size; the size of read and write pools may differ for (\subref{fig:overlapIO}) and (\subref{fig:nooverlapIO}). 
        }
        }
}
\label{fig:Interference-Aware}
\vspace*{-3mm}
\end{figure*}

WiscSort contains four algorithmic innovations to achieve these design goals.  First, WiscSort utilizes \textbf{key-value separation} to reduce I/O amplification, avoid redundant reads, and trade more reads for fewer writes.  Second, WiscSort includes a \textbf{thread-pool controller} that determines the number of threads for reads and writes for the device for managing access concurrency.  Third, WiscSort introduces an \textbf{interference-aware scheduler} that ensures reads and writes are not overlapped to avoid interference.  Finally, WiscSort capitalizes on the fact that keys and values are separated to sort larger amounts of data in memory in a single pass without a merge phase (\textbf{OnePass}).

Similar to external merge sort, WiscSort has two phases: the run generation phase and the merge phase. Instead of creating runs of key-value pairs (as in external merge sort), WiscSort creates runs of key-pointer pairs which refer to the values in the original input files. We call these new runs {\em IndexMaps}. During run generation, multiple threads read disparate partitions of the input file into DRAM and create key-pointer pairs; multiple sorting threads then concurrently sort individual runs; finally, multiple write threads persist the runs as IndexMap files. In merge phase, all the IndexMap files are read concurrently and merged. The final sorted data is then persisted.

\textit{Compliance with BRAID model:} WiscSort reads only the keys from the device, leaving the values in place. As the value is not utilized to produce the ordered records, there is no motivation to read it to perform the sort. WiscSort reads only the keys, which is facilitated by the \textbf{\textit{(B)}} property. Splitting the records leads to reading keys at strided locations (non-sequential). Due to \textbf{\textit{(R)}}, there is a minimal performance impact. Additionally, as the values never follow the keys while sorting, the amount of data written during the run phase is vastly reduced, addressing the \textbf{\textit{(A)}} property.

WiscSort performs interference-aware scheduling of read and write operations addressing the \textbf{\textit{(I)}} property. At any given point, either reads or writes are issued, thereby avoiding the interference created by read-write operations. WiscSort uses the thread pool controller to appropriately size the pool for a given access type (reads or writes) on the device, achieving the \textbf{\textit{(D)}} property.

Lastly, WiscSort can sort in just one pass, bypassing the creation of IndexMap files when the keys and their pointers can fit in the DRAM. We call this version \textit{WiscSort OnePass}. If the total size of the keys and pointers does not fit in the memory, WiscSort, like external merge sort, performs the run and merge phases. We call this version \textit{WiscSort MergePass}.

\subsection{Key-Value Separation}

Maintaining values in the run files and reading and writing values when not used for sorting is one of many external merge sort performance pitfalls. WiscSort is motivated by a simple revelation that keys and values can now be separated because of the byte-level granularity offered by BRAID \textit{(B)} without massive amplification costs during reads and writes.
WiscSort reads only keys from the record in a strided fashion leading to non-sequential reads. Each key read has a pointer associated with it to represent the file offset of the record. We call this key-pointer combination an {\em index} and the list of key-pointers an {\em IndexMap}. The IndexMap is stored in the memory during sorting and is later persisted.

To understand the impact of splitting records, let us consider a simple example. Assume a dataset that comprises records with a 10-byte key and a 90-byte value. 
Traditional external merge sort will read 100 bytes (10 + 90), as it focuses on harnessing sequential read performance; WiscSort only reads the 10B key, resulting in a $10x$ reduction in read I/O traffic. Moreover, the run files are read during the merge phase again, leading to another 10x reduction (or more if multiple merge phases are required). Similarly, there is also a significant reduction in the write traffic also. Instead of the value, only a 5-byte\footnote{5B represents $2^{40}$ ($\sim$1 trillion) record offsets, irrespective of the size of the record. 
8B can be used if larger dataset is required, resulting in a 5x write traffic reduction.} pointer is persisted along with the key. In the example, there will be a $\sim$7x reduction in the write traffic. As the writes are slower than reads (Property $A$), the write traffic reduction improves the performance. 

WiscSort has $2N(V-P)$ less read and write traffic compared to external merge sort in the worst case (MergePass) and $2N(K+V)$ bytes traffic reduction in the best case (OnePass), where $N$ is the number of records and $K, V, P$ are the key, value and pointer sizes.


\subsection{Thread-Pool Controller}
\label{tpool-control}
Traditional high-performance applications want to maximize CPU utilization, so they tend to overlap all operations when possible. In the no-synchronization concurrency model shown in Fig \ref{fig:nosync}, each thread in the pool repeatedly reads some data, sorts it, and writes it to a file. However, in this approach, there is no way to \textit{control} the number of concurrent threads performing a particular action, and straggler threads may overlap read write operations, causing minor read-write interference. In WiscSort, the thread-pool controller determines the thread pool size to be used for a particular operation (read, sort, and write) as shown in Fig \ref{fig:overlapIO} and \ref{fig:nooverlapIO}.

Given that a wide range of BRAID devices will exist in the future, deciding the number of requests to be sent concurrently is a non-trivial task. In our system, a microbenchmark determines the device's peak bandwidth capabilities and scaling behavior. The controller then utilizes this information at run time to determine the thread pool sizes. For example, the pool size of reads is much larger than that of writes (Figure \ref{fig:nooverlapIO}). 
In future Linux versions, one can directly gather the device performance data from Heterogeneous Memory Attribute (HMAT) tables \cite{hmat}.

\begin{figure*}
    \centering
    \includegraphics[scale=0.46]{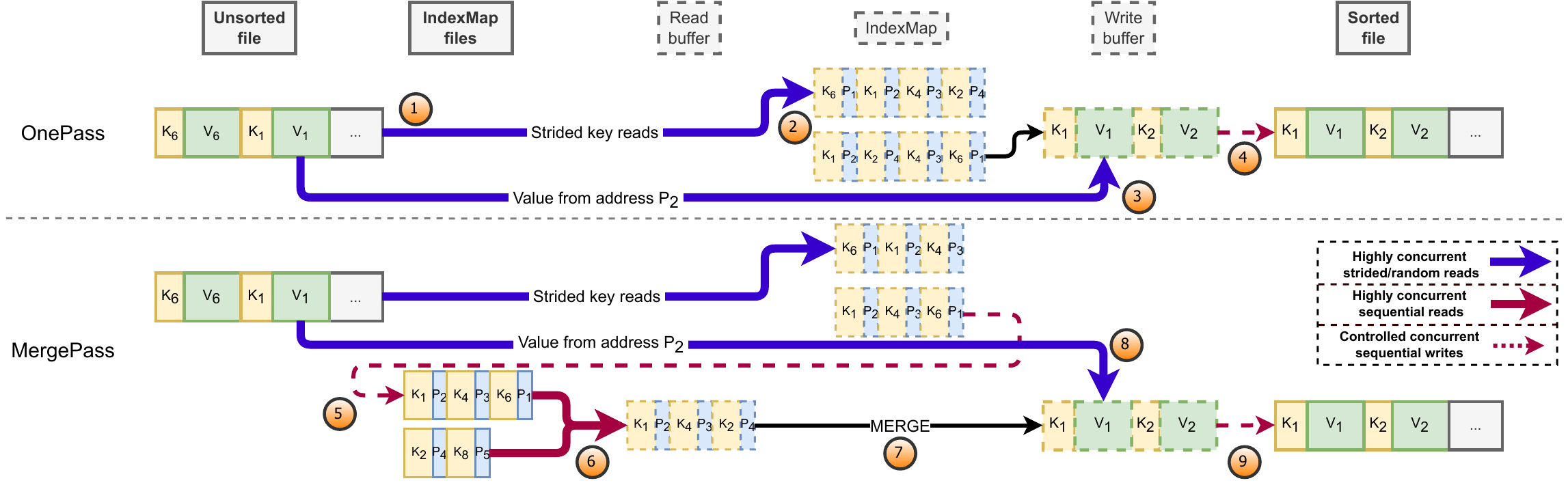}
    \caption{Data flow diagram of WiscSort OnePass and MergePass.}
    \begin{FlushLeft}
	{\texttt{K, V, and P }\small represent a Key, Value, and Pointer. Pointer is the offset at which the corresponding value exists in the input file. Solid boxes indicate data persisted on BRAID and the dashed boxes show data in DRAM. There are six different stages a key can be in, as shown at the top of the figure. Solid arrows are reads, and dashed arrows are writes. The red color represents sequential accesses and the blue non-sequential. The thickness of the arrow is proportional to the concurrency in the corresponding operation.}
    \end{FlushLeft}
    \label{fig:flow}
    \vspace{-3mm}
\end{figure*}

\subsection{Interference-Aware Scheduling}
\label{IAS}
As observed in many prior works  \cite{syseval-nvm, olap-nvm, emperical-guide-optane}, writes do not scale well on BRAID, and, therefore, one may be motivated to overlap reads and writes as shown in Figure \ref{fig:overlapIO}. However, BRAID's performance interference between reads and writes (($I$)) nullifies the benefit gained by overlapping them. Thus, it is important to isolate read and write operations.

WiscSort relies on interference-aware scheduling, where only read or write operations are issued at any given time, as shown in Figure \ref{fig:nooverlapIO}. The majority of the read operations occur at the start of the run phase and merge phase, while the write operations occur towards the end of the phases. While the data is being read in the run phase, WiscSort uses a write buffer to temporarily store the sorted data in the memory and periodically flushes the write buffer while stalling the reads. Similarly, the values gathered during the merge phase are isolated from the writes to the device through a write buffer. These buffers act as a logical barrier between different kinds of operations.  The write buffers are essential for adding a control point to separately size the number of reader and writer threads and to avoid any interference between these operations.  



\subsection{More Keys in One Pass}
Traditional sorting algorithms sort keys in a single pass (without writing intermediate results) when all keys and values fit into main memory, but must perform a second pass merge phase when the keys and values exceed memory capacity.
In contrast, splitting keys from values (via the IndexMap) reduces the memory footprint of WiscSort, and enables it to sort keys in a single pass when keys and pointers fit in main memory.

\revision{ In the one-pass version of WiscSort, the IndexMap is concurrently loaded into memory through strided key reads and then sorted in-place.
If we ignored the $(I)$ property, the values could be moved directly from the input file to the output file. However, WiscSort-OnePass performs thread-pool and interference-aware scheduling with a write buffer to optimize performance.}
\subsection{Algorithm}
\label{sec:algo}
\revision{
Fig \ref{fig:flow} represents the operation of WiscSort, including how data flows and state changes across steps for fixed-size records.
Steps 1-2 are performed regardless of whether one or multiple passes will be needed.
Section \ref{variable_len_algo} will describe the minor changes required to handle variable length records.}

\revision{
\textbf{\circled{1} \textit{ RUN read:}} For a given input file, WiscSort determines the maximum IndexMap size that can fit in memory while aligned to the input file size.
Consecutively the appropriate portion of the input file is evenly partitioned amongst the reader threads.
Each reader thread performs a strided read of the keys in its partition to the unsorted IndexMap, generating record-id pointers on the fly to reference the offset of the value in the original input file. Since the record sizes are fixed and contiguous (Sec \ref{assumptions}), each pointer is a hex address, calculated as $(start\_address + record\_id * record\_size)$.}

\textbf{\circled{2} \textit{RUN sort:}} Once all threads finish reading, a concurrent in-place sample sort is performed on the IndexMap.  If the IndexMap fits into DRAM, WiscSort will perform just one pass as in steps 3-4; otherwise, WiscSort performs two passes as in steps 5-9.
\vspace{-2mm}
\subsubsection{\textbf{WiscSort OnePass}}\hfill

\textbf{\circled{3} \textit{RECORD read:}} The sorted IndexMap is divided across a predetermined number of read threads provided by the thread pool controller; each thread performs a random read for the value from the unsorted file and places it into a write buffer. 

\textbf{\circled{4} \textit{RUN write:}} Once the write buffer is full, it is written sequentially to the output sorted file. The write buffer enables interference-aware scheduling; however, without it, WiscSort would still be faster than external merge sort simply because of finishing the entire sorting in a single pass.
\vspace{-2mm}
\subsubsection{\textbf{WiscSort MergePass}}\hfill\\\
As seen in \textbf{Steps 1 and 2}, the keys are gathered and sorted in chunks equal to that of the IndexMap size.

\textbf{\circled{5} \textit{RUN write:}} Since a single IndexMap does not contain all the keys of the input file, it is temporarily written to a file on BRAID in a sequential and concurrent manner; this does not require an output buffer.
\textbf{Steps (1, 2, and 5)} are repeated until all the keys of the input file are read, where a set of sorted IndexMap files are generated. This marks the end of the run phase.

\revision{
\textbf{\circled{6} \textit{MERGE read:}} In the merge phase, the read buffer is split evenly amongst the number of IndexMap files. Reader threads then sequentially load a chunk of each IndexMap to its appropriate area in the read buffer. 
Once a set of keys from all the IndexMaps fill the read buffer, a set of cursor pointers indicates the current and the end for the space allocated to the IndexMap file. The current cursors indicate the keys to be compared across the IndexMap files, and the end cursor pointer points to the last key in the space allocated for that IndexMap. 
These pointers are reset every time a new set of keys is read from the IndexMap file to its space in the read buffer.}

\textbf{\circled{7} \textit{MERGE other:}} WiscSort finds the minimum of the keys pointed to by the current pointers. The minimum key is then enqueued to an offset queue that maintains a list of pointers whose values must be read into the write buffer. WiscSort does not fetch the value after finding each min key because single thread random read bandwidth is poor.

\textbf{\circled{8} \textit{RECORD read:}} The size of the offset queue is determined by the size of the write buffer. Once the offset queue is full, WiscSort performs concurrent random reads of records to retrieve the values from the input file and update the write buffer.

\textbf{\circled{9} \textit{MERGE write:}} Once the write buffer is full, it is  sequentially written to the output file. The write thread-pool is controlled as per the device characteristics.
If any current pointer reaches the end pointer, WiscSort will read the next set of keys from the respective IndexMap file to its allocated space in the read buffer. 
If all the keys are already read from that IndexMap file, the read buffer space allotted to this IndexMap will be transferred to a neighboring IndexMaps evenly and the number of keys compared to find the minimum key will be reduced by one. 
The pointers are also updated accordingly when keys are read for the neighboring IndexMap.

Finally, if only one IndexMap file remains, it is loaded completely to the read buffer. If WiscSort has finished traversing of all the IndexMap files and the write buffer is still not full, the partially full buffer is flushed to the byte-addressable storage concurrently, marking the end of the merge phase. Throughout all of these steps WiscSort carefully ensures that the reads and writes never overlap through interference-aware scheduling.

\vspace{-2mm}
\subsubsection{\textbf{WiscSort for variable length values.}}\hfill\\\
\label{variable_len_algo}
Sorting the variable length records with fixed size keys (KLV - Sec \ref{assumptions}) requires only two changes, the IndexMap layout, and the random read processing. The IndexMap file will now contain one additional attribute, the length of the value. So an IndexMap is now a list of \texttt{(<key, pointer, vlength>)} entries, and \texttt{pointer} now points to the byte-offset of the corresponding value. The following steps indicate the changes to the previously described algorithm,

\textbf{\circled{1} \textit{RUN read:}}
Since the key byte offsets are unknown, concurrently reading the key and vlength is impossible. A single reader thread must serially read the key + vlength of each record to determine the next address to read from. The next key to be read is determined by appending the vlength to the existing key byte offset. Hence, the IndexMap file must be loaded serially in the RUN phase when dealing with the KLV format. This is restriction is shared by other sorting algorithms as well.

\textbf{\circled{3} \& \circled{8} \textit{RECORD read:}} The offset queue used to perform concurrent random reads now maintains vlength along with the list of pointers (sorted) that must be read to the write buffer. Once this queue is full, the thread-pool controller evenly partitions the queue among an optimal set of reader threads. Each reader thread now reads the value of vlength size from the input file to the write buffer. Once the write buffer is full, it is written to the BRAID device.

\subsection{Implementation}
To saturate the BRAID bandwidth, we must determine the right number of read and write threads and access granularity. The thread-pool controller relies on this information to decide the pool sizes. Thus, we developed a microbenchmark suite to characterize the device's performance. In our setup ( Sec \ref{setup}), read bandwidth scales up to 16 threads ($\#$physical cores) and saturates after that. Therefore, our implementation uses 16 to 32 threads (sequential and random) for reading data to the read buffers and 5 threads for writing from the write buffer, since writes do not scale with more concurrency.\looseness=-1

External merge sort slightly benefits from a large read buffer during the merge phase, as the number of times the read operations are to be performed is reduced. In the case of WiscSort, the read buffer sizes determine the number of passes required. If the read buffer is small and the entire IndexMap does not fit into the buffer, then MergePass will have to be used. So when possible, larger read buffers and smaller write buffers are preferred in WiscSort. The size of the write buffer has no performance significance.

To perform in-place sort of keys and index/pointers, we employ the state-of-the-art sorting implementation $IPS^4o$ \cite{IPS4o}. The concurrency is implemented using C++'s standard threading module \texttt{std::thread}. The synchronization between threads performing an operation is achieved through a condition variable indicating if all the threads have finished their portion of the work. We employ concurrent operations whenever possible; for example, loading keys from the read buffer to the key array \texttt{(<key,read\_buff\_ptr>)} and moving keys from the key array and values from the read buffer to the write buffer are all performed concurrently. Finally, for optimized I/O accesses, we use \texttt{AVX 512} non-temporal stores followed by a \texttt{clflushopt} for writes and \texttt{AVX 256} instructions for reads.

\section{Evaluation}
In this section, we evaluate the benefits of the design choices made in WiscSort. We compare WiscSort to well-established sorting algorithms on standard sorting benchmarks and show how WiscSort effectively utilizes a real byte-addressable storage device -- Intel Optane DC PMEM. Finally, we show the effectiveness of our techniques on emulated BRAID devices with varying properties.

\label{setup}
All the experiments are run on a test machine with one Intel(R) Xeon Gold 5218 $2^{nd}$ Gen CPU with scaling governor set to \texttt{performance}. There are two 16GB @2400MHz DRAM and four 128GB Intel Optane DC PMEM 100 @2666MHz placed on six distinct memory channels, as advised by Intel \cite{intel-pmem-populate}. The operating system is 64-bit Linux 5.0, and the PMEM devices are configured to App Direct mode with \texttt{fsdax} namespace and ext4 as the file system.

To evaluate the effect of Key-Value Separation, Thread-Pool Controller, and Interference-Aware Scheduling,
we ask:
\begin{enumerate}[leftmargin=*]
    \item How does WiscSort perform on popular application benchmarks (e.g., sortbenchmark)? Does WiscSort utilize the BRAID device bandwidth effectively?
    \item What is the benefit of concurrency optimizations? How does PMSort compare against WiscSort?
    \item What is the benefit of key-value separation? How does the benefit vary with different key:value ratios? Should sequential reads be preferred over random reads for all key-value sizes when generating the IndexMap from the input?
    \item How do WiscSort and other sorting methods perform on future BRAID devices (e.g., CXL) with different characteristics? 
\end{enumerate}

We answer the above questions using a series of microbenchmarks and industry-standard sorting benchmarks, and employ well-established techniques to emulate future BRAID devices \cite{first-gen-cxl}.
\\\textbf{We find that:} 
\begin{enumerate}[leftmargin=*]
    \item WiscSort OnePass is $3x$, and MergePass is $2x$ better than concurrent external merge sort on the sortbenchmark. Moreover, WiscSort saturates the device bandwidth for a given operation, demonstrating the benefits of conforming to the \textit{BRAID} model.
    \item Being device concurrency aware provides up to 50\% improvement in total time of WiscSort compared to I/O overlapping WiscSort counterparts.
    Also, WiscSort OnePass is $7x$ and \\ MergePass is $4x$ faster than PMSort.
    \item WiscSort offers more improvements over external merge sort with larger $V:K$ ratios. WiscSort OnePass outperforms external merge sort regardless of the $V:K$ ratios, and MergePass outperforms external merge sort when $V:K > 1$. In addition, loading IndexMap via strided reads is always beneficial regardless of $V:K$ ratios.
    \item Experiments on future CXL BRAID devices show external merge sort and WiscSort OnePass are most favorable amongst others on devices with poor random-read performance, large asymmetric read-write costs, and symmetric costs.
\end{enumerate}


\subsection{SortBenchmark}

\begin{figure}[!t]
    \begin{minipage}{0.48\textwidth}
    \centering
    \includegraphics[width=0.82\textwidth]{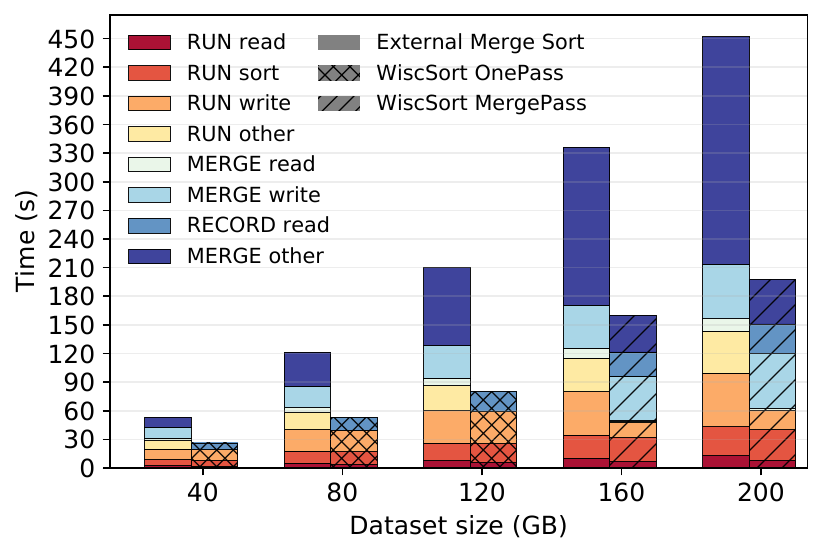}
    \vspace*{-3mm}
    \caption{
    \revision{
    WiscSort and external merge sort performance on sortbenchmark workload. \textmd{Since the key values sizes are fixed, the speedup between WiscSort and external merge sort is consistent for varying file sizes. 
    }
    }}
    \label{fig:graysort}
    \end{minipage}
\end{figure}

\begin{figure*}
\centering
\begin{subfigure}{.482\textwidth}
    \includegraphics[width=0.82\textwidth, right]{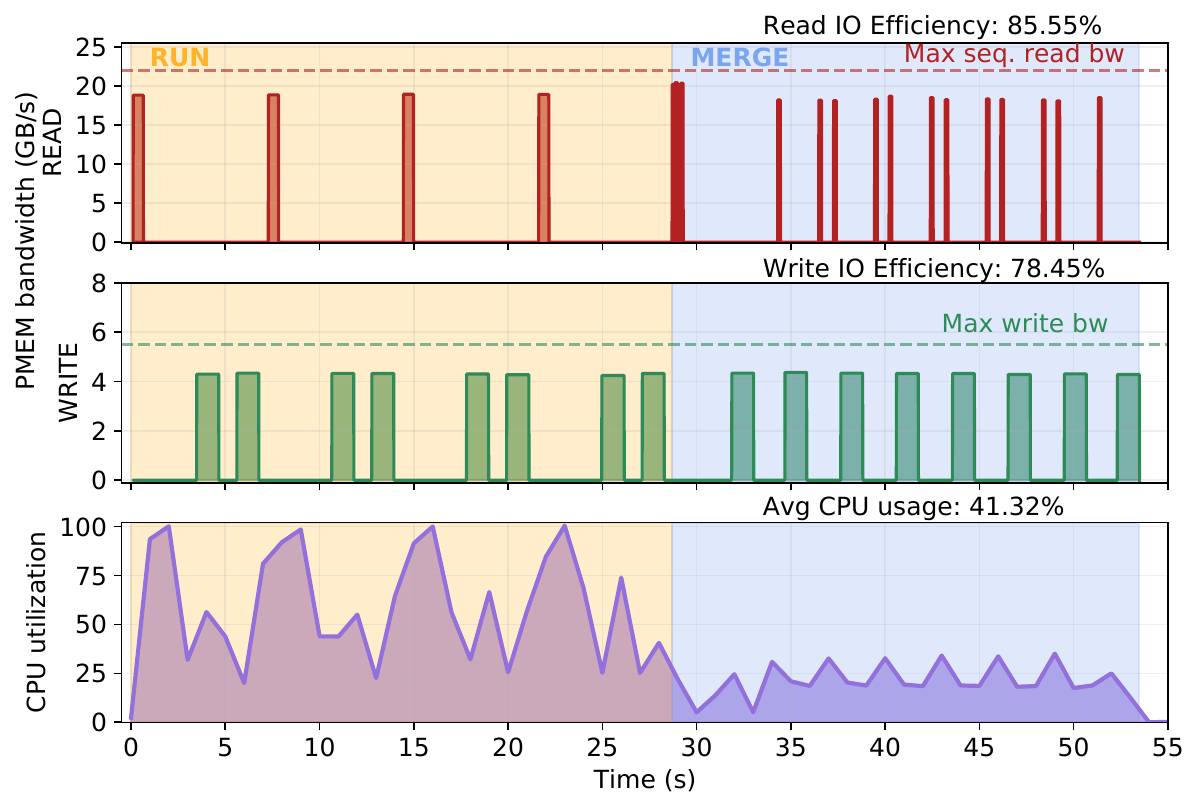}
    \caption{External merge sort. \textmd{10GB read buffer, 5 GB write buffer}}
    \label{fig:ems-resource}
\end{subfigure}%
\begin{subfigure}{.46\textwidth}
    \includegraphics[width=0.82\textwidth, left]{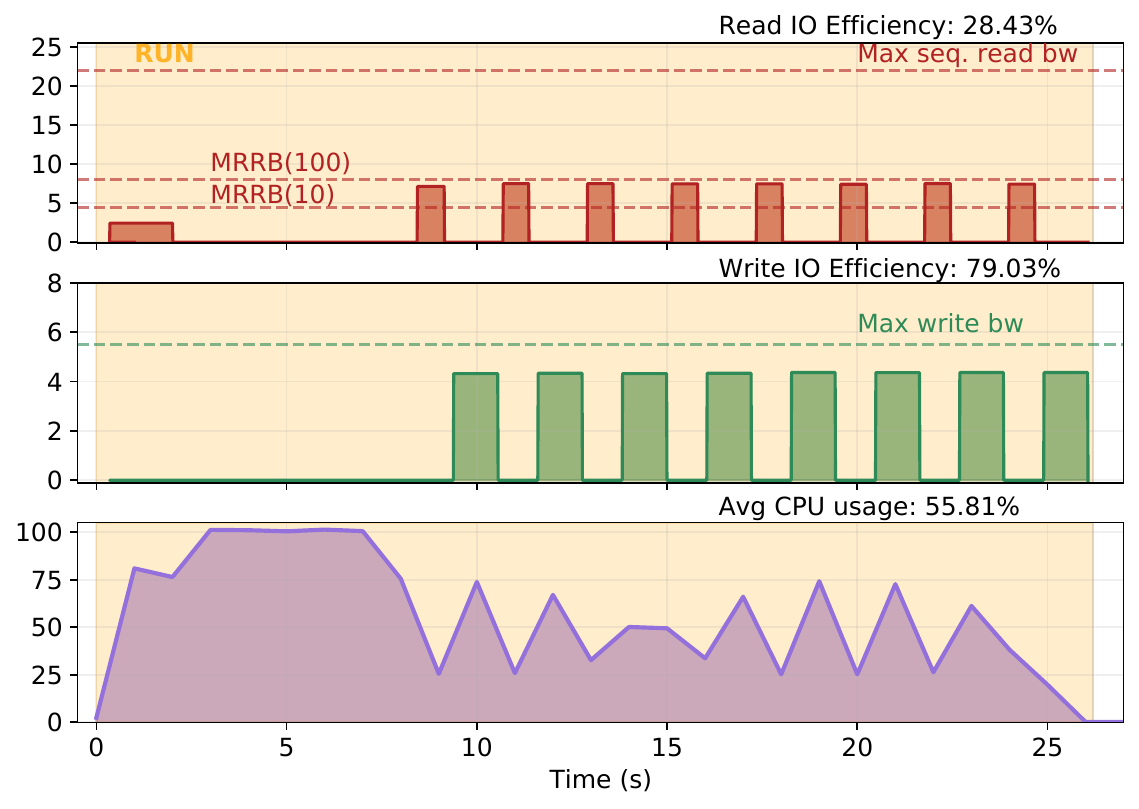}
    \caption{WiscSort OnePass. \textmd{With 5 GB write buffer}}
    \label{fig:onepass-resource}
\end{subfigure}
\vspace{-3mm}
\caption{Resource usage of external merge sort $(I + D)$  and WiscSort OnePass $(B + R + A + I + D)$ for sorting a 40 GB file. \textmd{ The dotted lines represent the peak bandwidth possible. As reported by our microbenchmarks, the Max Random-Read Bandwidth(\textit{MRRB}) changes with access size.
I/O efficiency compares actual time to ideal time for data operation.
Ideal time = operation size / peak bandwidth. 
for example, the ideal time to read 20 GB on our setup is 0.90s (read size / max read bandwidth).
}}
\label{fig:resource-40}
\end{figure*}

The sortbenchmark, first introduced in AlphaSort \cite{alphasort}, is the de facto industry standard for stress testing I/O architectures \cite{mapreduce, tencentsort, nowsort, sparksort}. We evaluate WiscSort on sortbenchmark workloads as a representation of application workloads. The benchmark is to sort binary records with 10B keys and 90B values. The input file to be read and the generated output file must be placed on BRAID (e.g., PMEM). The keys are of uniformly random distribution, and the output file must be a permutation of the input file, sorted in key ascending order.
As shown in Figure \ref{fig:graysort}, WiscSort consistently outperforms (2-3$x$) a competitive external merge sort implementation (using thread-pool controller and interference aware scheduling) for all sortbenchmark dataset sizes due to WiscSort's design choices (conforming to BRAID properties). The legend in the figure maps to the operations described in Sec \ref{sec:algo}.

WiscSort can sort up to 200 GB of input data on 448 GB of usable PMEM using a 5-byte pointer. The 200GB dataset benchmark is the largest workload we can run with our PMEM capacity. The benchmark requires 30 GB for IndexMap files and another 200 GB for the output file.
When evaluating WiscSort MergePass, we limit the available DRAM capacity (32 GB) to 20 GB so that the IndexMaps of input files larger than 140GB do not fit entirely in the DRAM. The external merge sort uses a 10 GB read buffer and 5 GB write buffer. WiscSort only uses a 5 GB write buffer.
The buffer size choice has no effect on either sorting system.

\begin{figure*}
\vspace{-4mm}
\centering
\begin{subfigure}{.483\textwidth}
  \centering
    \includegraphics[width=0.82\textwidth, right]{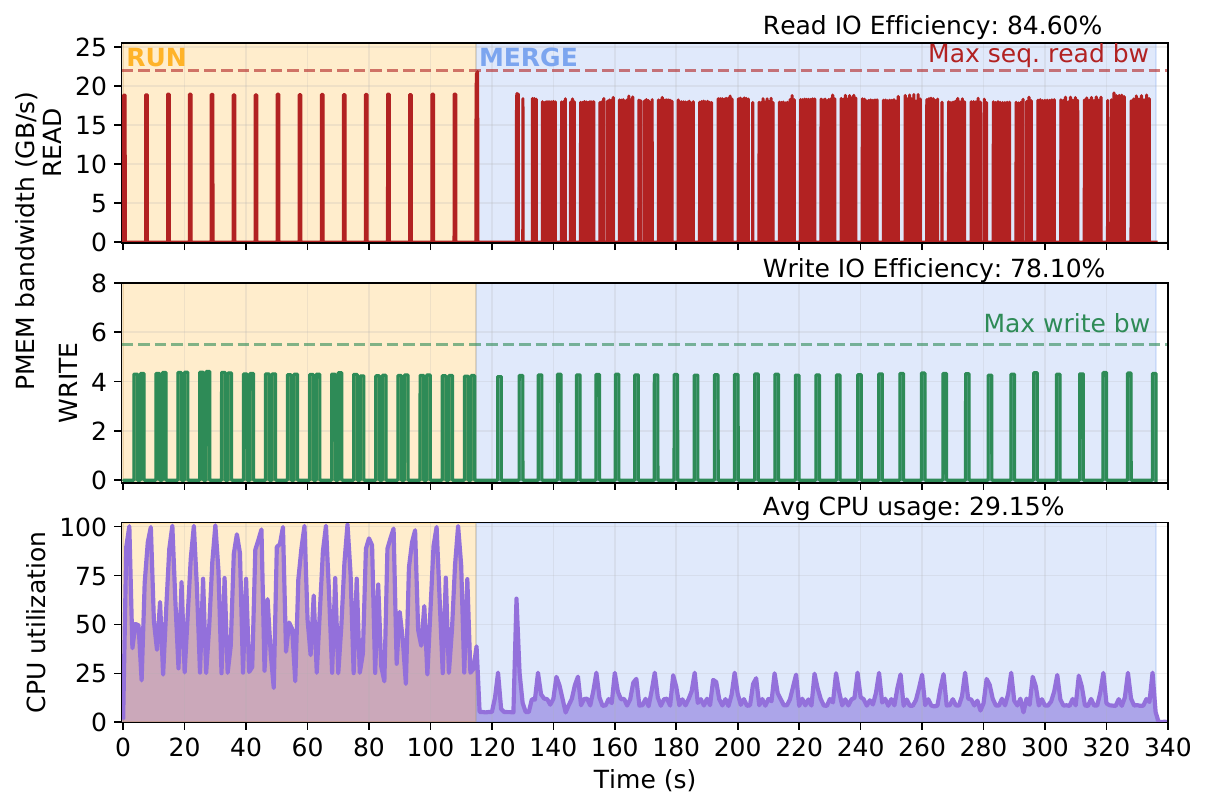}
    \caption{External merge sort. \textmd{10 GB read buffer, 5 GB write buffer}}
    \label{fig:ems-resource-160}
\end{subfigure}%
\begin{subfigure}{.46\textwidth}
  \centering
    \includegraphics[width=0.82\textwidth, left]{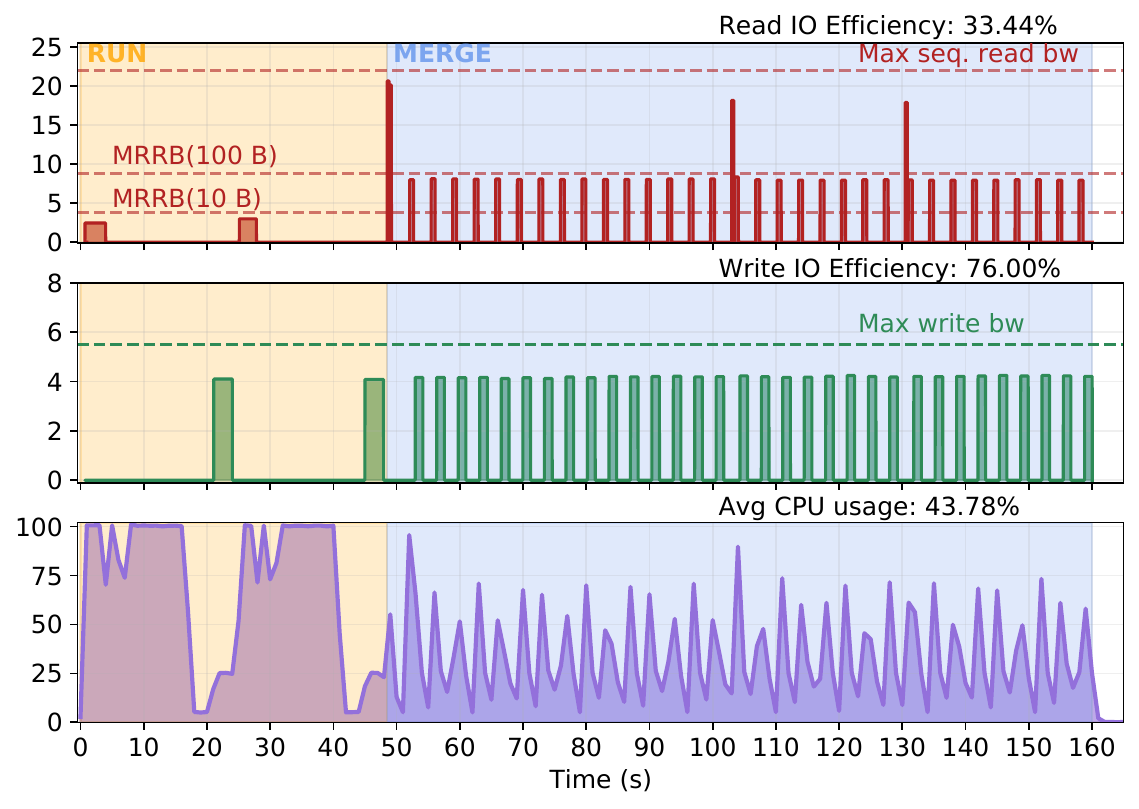}
    \caption{WiscSort MergePass. \textmd{12 GB read buffer, 5 GB write buffer}}
    \label{fig:mergepass-resource}
\end{subfigure}
\vspace{-3mm}
\caption{Resource usage of external merge sort  $(I + D)$  and WiscSort MergePass $(B + R + A + I + D)$ for sorting a 160 GB file.}
\label{fig:resource-160}
\vspace{-3mm}
\end{figure*}

\textbf{WiscSort OnePass} is up to $3x$ faster than the external merge sort for the 40GB/80GB/120GB datasets. This is due to the $50\%$ reduction in read/write traffic and the avoidance of the merge phase computation. The external merge sort is at least $25\%$ slower than WiscSort during RUN read, with larger datasets (more value read traffic), the gap can reach up to $60\%$. 
The total in-memory sort times (RUN sort) are the same as the key array, and IndexMap is of similar size. In the 40 GB case, the RUN other in external merge sort accounts for $12\%$ of total time; this includes the time taken to copy records from the read buffer to the key array and values from the read buffer to the output buffer concurrently. 
In contrast, WiscSort does not have the RUN other overheads because it does not have related operations, further reducing computation overhead.
The total write time during the run phase is the same between the external merge sort and OnePass since external merge sort has to write all the key values to the run files, whereas WiscSort OnePass writes all the key values in the final sorted order to the output file.

Figure \ref{fig:resource-40} and \ref{fig:resource-160} show the PMEM bandwidth and CPU resource usage of external merge sort, OnePass, and MergePass.
As shown in Figure \ref{fig:resource-40}, WiscSort consumes less bandwidth than external merge sort due to its strided key reads and random reads of values.
Figure \ref{fig:onepass-resource} shows the repeated random reads made to fetch the records from the input file pointed by the indexes in the sorted IndexMap file (RECORD read). The strided key reads have lower bandwidth due to the smaller accesses performed (10B key compared to 100B records).
Although reading the entire dataset sequentially to memory (MERGE read) is faster than RECORD read, OnePass avoids all other merge phase operations warranting the use of slower random reads ($B+A+R$ property). 
The thread-pool controller sizes the pool appropriately to ensure all I/O operations (sequential and random reads, writes) perform at peak bandwidth as shown in Figure \ref{fig:resource-40} \& \ref{fig:resource-160}.

\textbf{WiscSort MergePass} is up to $2x$ faster than external merge sort for the 160GB/200GB datasets, due to the total reduction in read/write traffic (42.5\%). Like OnePass, MergePass takes (considerably) less time to load IndexMap files, comparing to external merge sort. 
Unlike OnePass, MergePass persists IndexMap files on the device (as in Figure \ref{fig:mergepass-resource}). 
During the merge phase, MergePass saturates device read bandwidth due to its sequential loads of the portion of the IndexMap file; in MergePass, there are fewer IndexMap loading reads (compared to external merge sort), since more keys can fit in DRAM due to key/value separation. 
As a result, for example, with the 160 GB dataset, WiscSort MERGE read time is $7x$ smaller than that of the external merge sort. In the merge phase, WiscSort MergePass generates random record value reads once keys are ordered after merge. Due to interference-aware scheduling, in WiscSort MergePass, no two types of I/O operations overlap.

MERGE writes dominate the overall time of WiscSort MergePass. The total write time of external merge sort is $2$x of the total write time of WiscSort OnePass and $~1.5$x that of WiscSort MergePass, satisfying the $(A)$ property.
MERGE other time indicates operations other than reads and writes in the merge phase. 
In external merge sort, a single thread finds the minimum between keys from each run file and copies the record from the read buffer to the write buffer. This cannot be made concurrent since all the RUN files are merged in a single merge phase. On the other hand, the WiscSort MergePass performs concurrent copies of records to the output buffer directly since the read offsets are accumulated and submitted at once, as depicted by the better CPU utilization in Figure \ref{fig:resource-160}. A similar optimization cannot be applied to external merge sort because the records at input buffer offsets can potentially change before they are concurrently copied to the output buffer.

Overall, we show that WiscSort maximizes random-read bandwidth $(R)$, reduces the amount of writes $(A)$, takes advantage of the byte addressability $(B)$ while maximizing CPU utilization, and is aware of the concurrency constraints of the device $(I + D)$, thus making it a \textit{BRAID compliant} algorithm. On the other hand, our implementation of external merge sort is only aware of $(I + D)$, making it a \textit{non BRAID compliant} system. 

\vspace{-3mm}
\subsection{Concurrency \& Interference Optimizations}

\begin{figure*}
    \captionsetup{justification=centering}
    \centering
    \begin{minipage}{0.31\textwidth}
        \centering
        \includegraphics[width=0.86\textwidth]{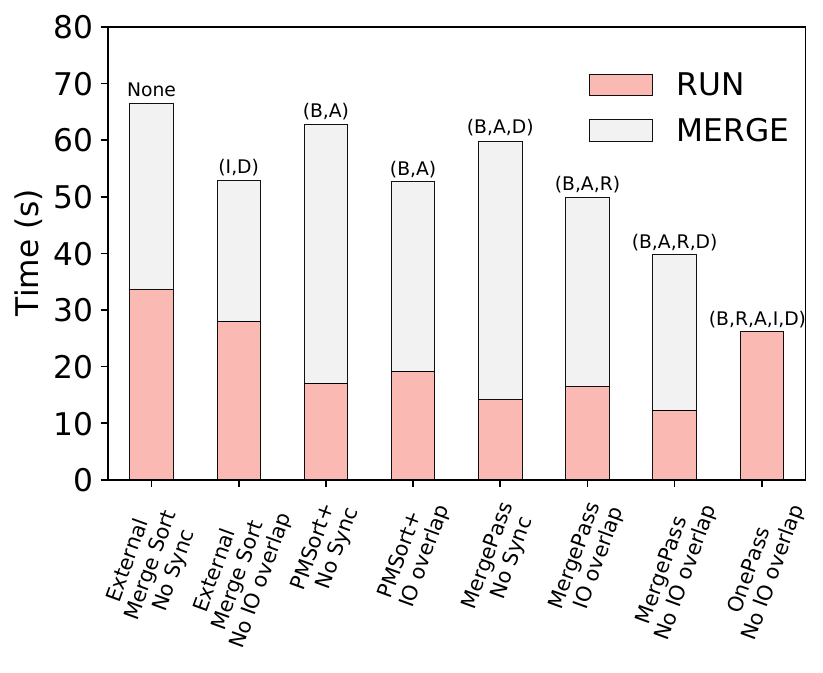}
        \vspace{-5mm}
        \caption{
        \revision{
        Sorting systems using different concurrency models and the BRAID properties they fulfill. \textmd{Sorting 400M records of 10B K: 90B V each.
        }}
        }
        \label{fig:micro-concur}
    \end{minipage}
    \begin{minipage}{0.32\textwidth}
        \centering
        \vspace{-3mm}
        \includegraphics[width=0.86\textwidth]{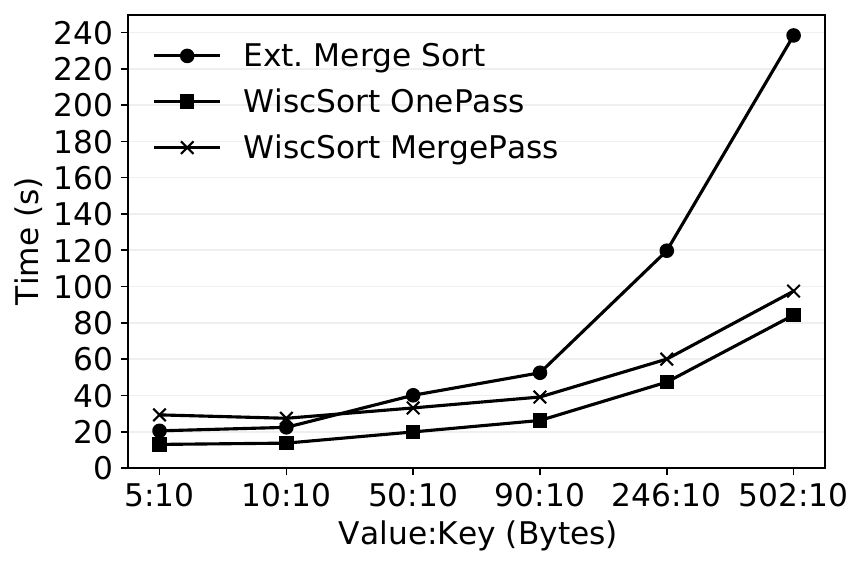}
        \caption{Key Value splitting benefits\\ for 400M records. \textmd{The key size is 10B, Pointer is 5B, and the value size varies.
        }}
        \label{fig:micro-varying-value}
    \end{minipage}
    \begin{minipage}{0.32\textwidth}
        \centering
        \vspace{-2.9mm}
        \includegraphics[width=0.86\textwidth]{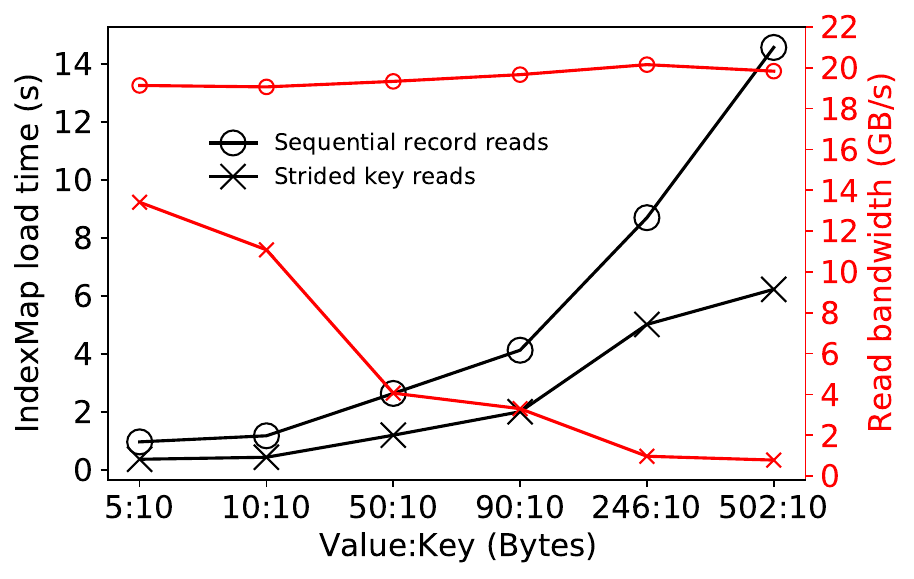}
        \caption{Load IndexMap by Strided vs. Sequential reads for 400M records. \textmd{The key size is 10B, and the value size varies.}}
        \label{fig:micro-load-access}
    \end{minipage}
\vspace{-3mm}
\end{figure*}

Figure \ref{fig:micro-concur} demonstrates the benefits of paying attention to the constrained concurrency and the read-write interference of a device. We compare multiple concurrency models (Figure \ref{fig:Interference-Aware}) of traditional external merge sort, PMSort, and WiscSort against each other. 
 
To differentiate the benefits gained by separating key and value, we compare the external merge sort No Sync (Figure \ref{fig:nosync}) and No I/O overlap (Figure \ref{fig:nooverlapIO}). During the merge phase of No Sync, we use a write buffer to make the output writes concurrent. The buffer enforces an order between read-write operations, which avoids interference, but it still suffers from write degradation due to uncontrolled pool sizes. The external merge sort No Sync has worse run time performance than any other multi-threaded sorting system, due to larger amounts of contention. Due to interference-aware scheduling and with thread-pool control, the No IO overlap performs 25.7\% faster than No Sync. Indicating that the thread-pool control and interference-aware scheduling can improve performance for any device-unaware algorithm.

The published version of PMSort separates keys and values, but does not perform strided gather of keys during the run phase; additionally, it avoids value writes at the end of the run phase. It ignores device concurrency characteristics, does not investigate effective resource utilization, and avoids random reads whenever possible. Since the PMSort codebase is not available, we implemented the single-threaded version as specified \cite{pmsort}; we also built PMSort multi-threaded versions based on traditional concurrency models (Figure \ref{fig:nosync} and \ref{fig:overlapIO}) calling it PMSort+. Our PMSort+ implementation queues the random read offsets in the merge phase, so that the value gathering can be concurrent as done in WiscSort. 

The merge phase of PMSort+ No Sync has no write buffer (unlike external merge sort No Sync). The values can be moved directly and concurrently from the input file to the output file once the offset vector is filled. This method, however, causes serious read-write interference coupled with poor write performance degradation, hence making No Sync 16\% slower than I/O overlap. 
During the merge phase of all I/O overlap systems, we maintain two write buffers and two offset vectors to ensure that the random reads and writes always overlap; this helps quantify the effect of thread-pool control alone -- the I/O overlap merge is 36\% faster than No Sync.

We implement all three concurrency variants of WiscSort to study the effects of BRAID compliance. The run phases of WiscSort, unlike PMSort, perform strided gather, thus reducing the overall run time in comparison. However, the merge phase time between the two remains the same. Because of this, both WiscSort I/O overlap and No Sync perform better than PMSort+ equivalents. Moreover, the MergePass no I/O overlap, which performs interference-aware scheduling and uses thread-pool controller, performs 33\% faster than the hypothetical best case of PMSort+ and \textbf{$\sim$4x} faster than the actual (single thread). If the IndexMap fits into memory, WiscSort OnePass is \textbf{\textbf{7x}} faster than single-threaded PMSort. The single-threaded WiscSort MergePass has no performance regression, although the single-thread random reads are bad. This slowdown is because of the redundant reads (of values) PMSort performs during the run phase and retaining only keys, causing two copies rather than one. Nevertheless, since intermediate writes are avoided, WiscSort OnePass will still be faster than PMSort.

Overall, we show that even a concurrency optimized external merge sort $(I + D)$ can outperform sorting systems with naive concurrency, even if they separate key and values $(B + A + R)$. Even amongst systems that comply with $(B + A + R)$, choosing the appropriate concurrency model can result in huge gains. We also highlight the importance of utilizing the concurrency feature of the device itself (PMSort single-thread vs. multi-thread). Finally, the considerable benefits of interference-aware scheduling and thread-pool control shown in Figure \ref{fig:micro-concur} will only grow with larger datasets.
 
\subsection{Random-Read Optimizations}
To better understand the key-value separation impact, we conduct experiments where we vary the Value:Key ratios. 
It's worth noting that most production KV workloads (like Facebook\cite{benchmark-rocksdb-fb}) have large values and small keys, making the key-value separation idea a good fit. 
In addition, we evaluate whether random/strided reads are more efficient than sequential reads for loading the IndexMap file with varying value-key ratios.
We generate the datasets required using a custom tool built on top of Gensort \cite{gensort}. 

As seen in Figure \ref{fig:micro-varying-value}, WiscSort performs up to 3x (OnePass) and 2x (MergePass) better than external merge sort $(I + D)$ when value sizes are larger than 90B. The improvement is due to the reduced traffic volume in WiscSort caused by the key-value separation. For WiscSort OnePass, the traffic reduction percentage is constant (50\%) regardless of the value size. For MergePass,  there is more traffic reduction with larger values; for example, MergePass has 48.5\% traffic reduction with 502B value but 37.5\% reduction with 50B value (compared to external merge sort).
Figure \ref{fig:micro-load-access} demonstrates the impact of such reduced traffic on IndexMap load time.

With values of medium sizes (i.e., 50B and 90B), Figure \ref{fig:micro-varying-value} tells a different story, where although there is a benefit of WiscSort MergePass, it is small in comparison. 
This decline is because of the lower traffic reduction and lesser impact of improved CPU usage of WiscSort during the merge phase. Since the key size is close to the value size, the benefits of splitting are small. Nevertheless, OnePass performs 2x better. The difference in loading the IndexMap file through sequential vs. random is also reduced. Similarly, PMSort performs sequential record reads to memory and then gathers keys and pointers from it; as shown in Figure \ref{fig:micro-load-access} it is up to $\sim$3x worse than performing strided gather of keys.

For cases where the value size is the same as the key size or even smaller, MergePass performs worse than the traditional external merge sort. This regression is because smaller record sizes exhibit poor random-read performance compared to sequential read on PMEM during value gathering in the merge phase. Even though strided reads have good read bandwidth because multiple records can fit the 256B cache line (17 15B records and 12 20B records), the random read during the merge phase does not make use of this.
However, external merge sort can reach peak read bandwidth even for small records because they are sequential. 
For values of 10B and smaller ($V: K < 1$), the write reduction will be minimal or none in the case of MergePass, thus splitting key-value unsuitable. However, OnePass does better because of property $(A)$. Due to the disparity of read-write costs, multiple concurrent writes are still costlier than a single concurrent random read.

Overall, we demonstrated that irrespective of the $V: K$ ratio separating them on a BRAID device is beneficial, given that the IndexMap file fits in the DRAM (OnePass). For larger $V: K$, MergePass benefits further from the write reduction $(A)$. Due to PMSort access patterns, it always underperforms in comparison to MergePass and OnePass for any $V: K$. This implication was derived from Figure \ref{fig:micro-load-access} where strided gather performs better than sequential reads irrespective of the $V: K$ ($R$ property).

\vspace{-6mm}
\revision{
\subsection{Background I/O interference effects}

Thus far, we demonstrated the benefits of interference and concurrency constraint awareness \textit{(I +D)}. However, at the OS level, the BRAID device will be utilized by multiple processes for which we do not have any control over the requests made. Moreover, in the context of a database, it may not be desirable to delay a write from a short transaction while a long read phase for another query is underway. While efficient BRAID utilization with multiple processes is not the focus of this work, we will demonstrate the robustness of WiscSort with varying degrees of I/O interference intensity. 

In Figure \ref{fig:interference-overhead}, we observe the slowdown of WiscSort and merge sort as they are subjected to read and write heavy background workloads with varying concurrency. Each thread/client executes a 4KiB read (Fig \ref{fig:read-int-over}) or write (Fig \ref{fig:write-int-over}) operation on a large file. None of the background clients share cores with themselves or the sorting workload. Although the size of accesses made by the background clients can impact the interference effect, we keep it constant at 4KiB to facilitate direct comparison between WiscSort and merge sort in the common case.

The impact of background read workloads on WiscSort and merge sort is minimal when compared to background write-heavy workloads. 
The primary sources of the slowdown in the case of background readers are random reads and writes (Fig \ref{fig:read-int-over}), whereas writes are the primary cause of the slowdown in the presence of background writers (Fig \ref{fig:write-int-over}). 
Random reader threads in the background affect sequential reads due to on-device prefetching overhead and limited device scratch memory.
Overlap of random read requests can lead to significant overhead, while sequential background reads have a negligible effect. For example, WiscSort experiences a slowdown of 45\% when executing 8 concurrent random reader threads, whereas merge sort's slowdown is only 25\%.

The presence of background write-heavy clients poses a significant challenge for the sorting workloads due to the poor scalability of writes on PMEM, leading to a substantial impact on write times. Nonetheless, WiscSort, which is BRAID-compliant, is always twice as fast as merge sort, regardless of write intensity. We observe a slowdown of up to 14x in both WiscSort and merge sort with eight overlapping writer threads. Random reads are considerably slower than sequential reads when overlapped with background writes. Overall, WiscSort outperforms merge sort for both read and write-heavy background workloads; however, a userspace IO scheduler is essential to manage multi-tenant workloads on BRAID devices.

\begin{figure}[!t]
\centering
\begin{subfigure}{.25\textwidth}
    \includegraphics[width=\textwidth, left]{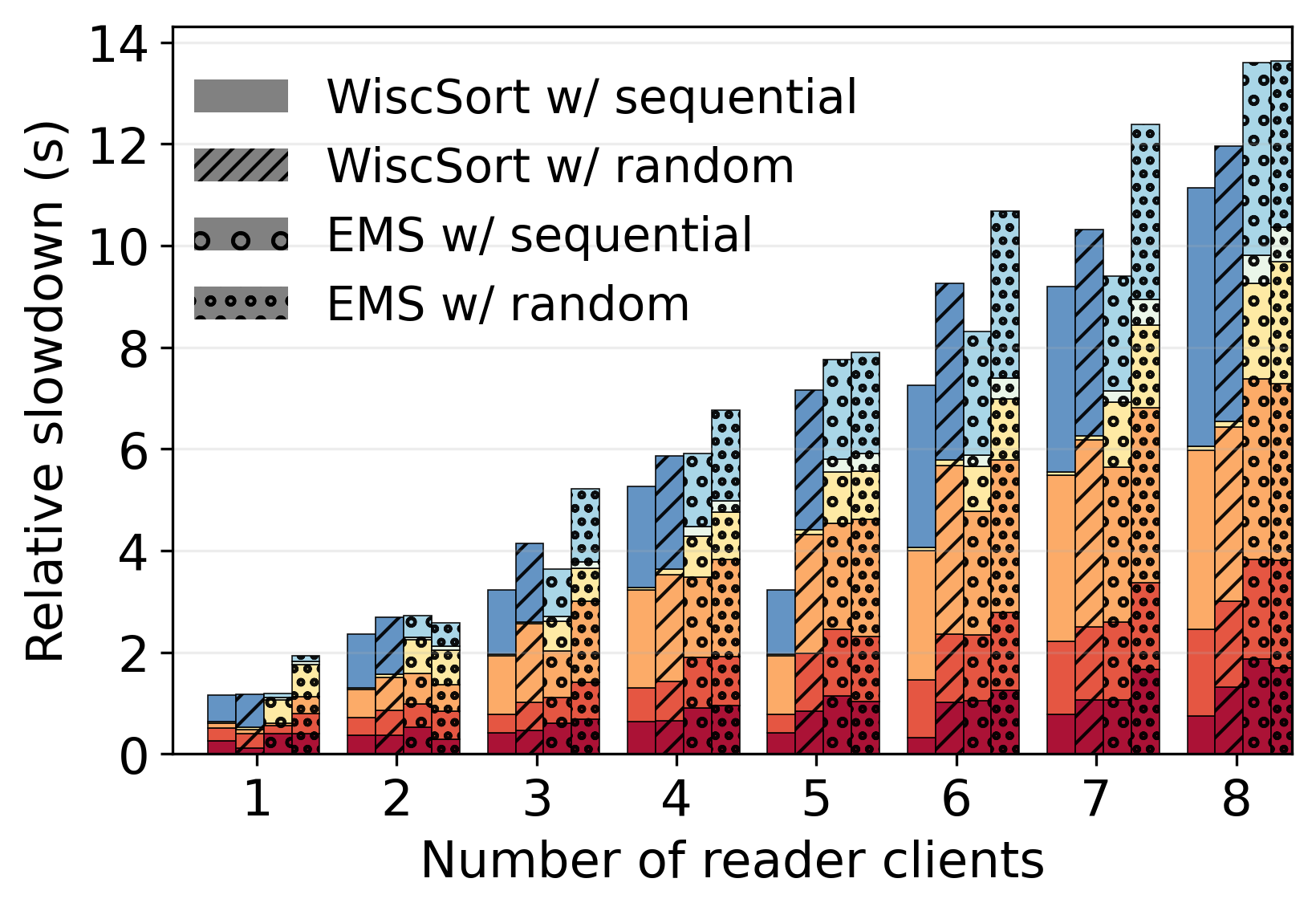}
    \caption{\revision{Read interference overhead}}
    \label{fig:read-int-over}
\end{subfigure}%
\begin{subfigure}{.245\textwidth}
  \centering
    \includegraphics[width=\textwidth, right]{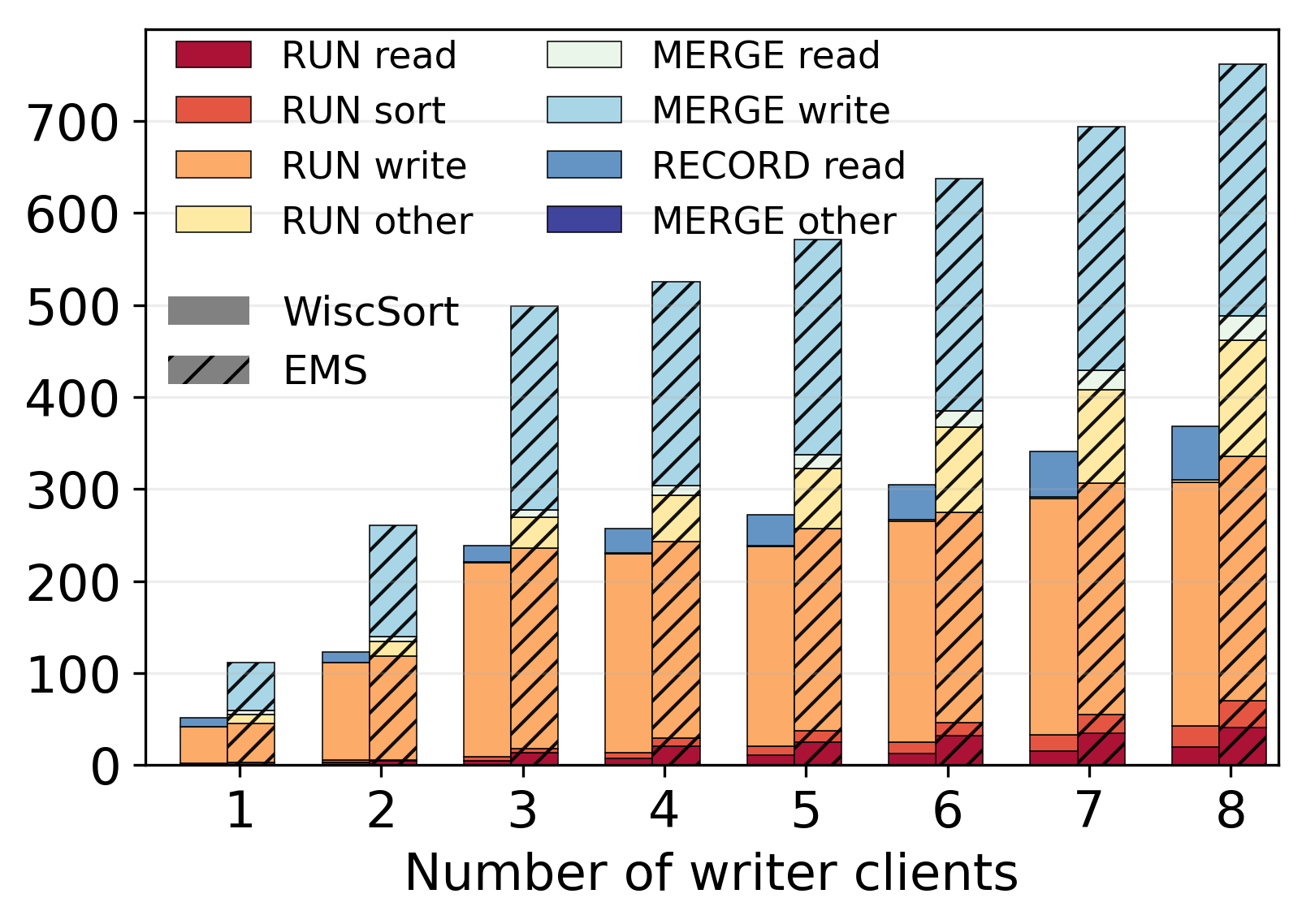}
    \caption{\revision{Write interference overhead}}
    \label{fig:write-int-over}
\end{subfigure}
\vspace{-3mm}
\caption{      
\revision{
            Interference effects of WiscSort OnePass and External Merge Sort (EMS) against multiple I/O intensive clients. 
                \textmd{ Sorting 400M records of 100B each. Each background thread performs 4KiB requests of read or writes to a different file on the device.}
        }}
\vspace{-25pt}
\label{fig:interference-overhead}
\end{figure}

}

\subsection{Other BRAID Devices}
\label{cxl-emulation}
Our previous experiments examined the WiscSort performance on Optane PMEM. However, future BAS devices may have other different performance characteristics. In the following experiments, we show the sensitivity of WiscSort and its optimizations on devices with various BRAID characteristics.
\revision{We also compare in-place sample sort \cite{IPS4o} (Sec \ref{samplesort}) and external merge sort on these new devices to determine what sorting technique works best on them.}

We emulate a single-socket system with a BAS device on a two-socket server by disabling all cores in one socket, while keeping its memory accessible from the other socket. This memory now mimics a CXL byte-addressable storage device. This is a well established emulation technique \cite{first-gen-cxl, enabling-cxl-db} that provides close to real CXL device performance \cite{samsung-cxl-expander}. The emulation test bed comprises two 20-core Intel Xeon Gold 5218R CPU with scaling governor set to \textit{performance} and hyperthreading disabled. The testbed constitutes 128 GB @2933MHZ DRAM split between two NUMA nodes equally. The operating system is 64-bit Linux 5.4, and the emulated BAS device is accessed through a file system interface created through \texttt{tmpfs}. Hence, the max BAS device capacity on this testbed is 64 GB. We inject delays through unoptimized \texttt{for} loops that busy loops until the desired wait time in nanoseconds is met. Each added delay is per cache line access (64B).

\subsubsection{\textbf{BD-Device:} Byte Addressable, Device Concurrency} \hfill \\
This device is inspired by traditional SSDs, which have \textit{symmetric} sequential read-write costs, but sequential reads are much faster than random reads. Thus this device does not exhibit the $(A)$ and $(R)$ properties of BRAID. BD-Device is an emulated byte-addressable 'disk' where random reads are $500ns$ slower than sequential reads.

Figure \ref{fig:cxl-ba} shows the performance of different sorting strategies; it especially highlights the pitfalls of a WiscSort-like design that relies completely on the random-read performance of a device. 
The sample sort performs direct in-place record movement to generate the sorted output without copying the entire input to local DRAM, hence paying a one-time cost of random access. 
External merge sort, as designed, performs best on BD-Device as it avoids random reads altogether.
In WiscSort, MergePass and OnePass pay a huge price due to their reliance on random reads during both the RUN phase (to generate the IndexMap file) and the merge phase (to gather values from the input file).
Showing us that external merge sort is the best sorting technique on the BD-Device of the BRAID model.

\subsubsection{\textbf{BRD-Device:} Byte Addressable, Higher Random Read, \\ Device Concurrency} \hfill \\
The proliferation of NVDIMM-N, larger on-device DRAM caching, and the introduction of new kinds of storage media (3DXpoint, XL-FLASH \cite{kioxia-cxl}) suggests that future BAS devices will have improved random-read bandwidth and write performance. One can expect DRAM-like random read and write performance from storage devices soon. BRD-Device is an emulated BAS with equal random read, sequential read, and write performance. Our emulation testbed does not need any modifications to emulate this device.

\begin{figure*}
\captionsetup{justification=centering}
\centering
\begin{subfigure}{.33\textwidth}
  \centering
    \includegraphics[width=0.88\textwidth]{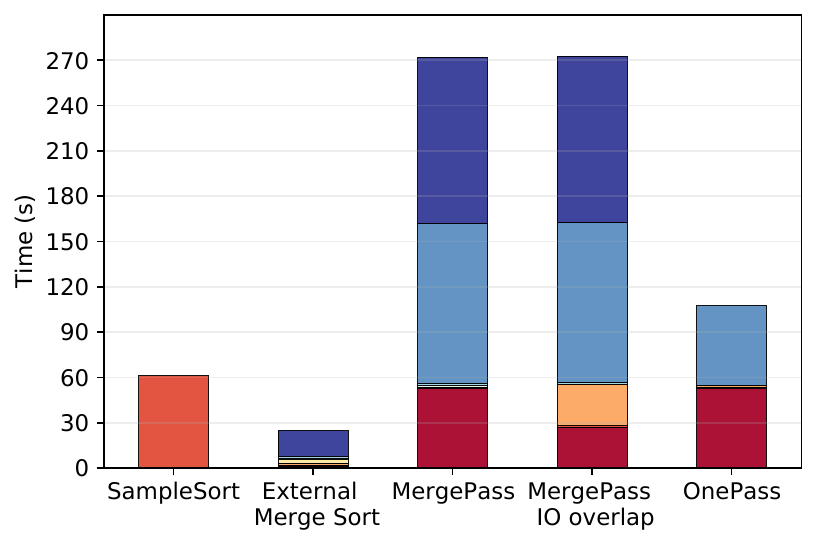}
    \caption{$BD$-Device. \textmd{Random reads are 500ns slower than sequential reads.}}
    \label{fig:cxl-ba}
\end{subfigure}%
\begin{subfigure}{.33\textwidth}
  \centering
    \includegraphics[width=0.88\textwidth]{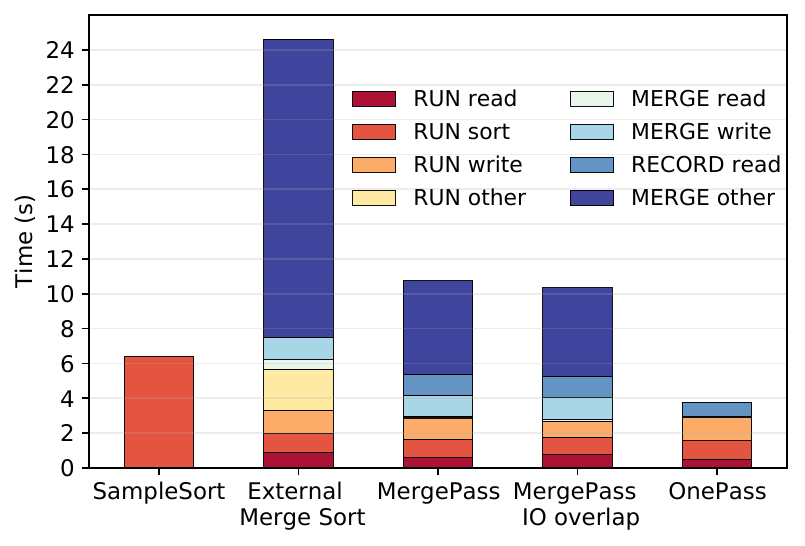}
    \caption{$BRD$-Device. \textmd{Random read is equal to sequential read.}}
    \label{fig:cxl-ba-RR}
\end{subfigure}
\begin{subfigure}{.33\textwidth}
  \centering
      \vspace{-6pt}
    \includegraphics[width=0.89\textwidth]{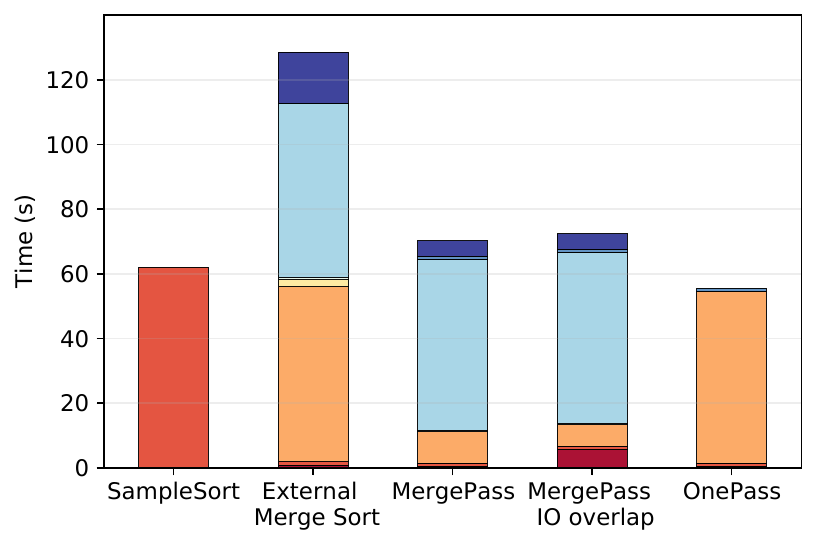}
    \caption{$BARD$-Device. \textmd{Writes 500 ns slower than reads.}}
    \label{fig:cxl-ba-RR-ARW}
\end{subfigure}
\vspace{-3mm}
\caption{Future BRAID device devices through CXL emulation. \textmd{Sorting 100M records with 10B Key and 90B value each.}}
\label{fig:cxl}
\vspace{-3mm}
\end{figure*}

As expected, Figure \ref{fig:cxl-ba-RR} shows that WiscSort OnePass performs the best among the sorting systems.
Since BRD-Device does not exhibit concurrency constraints such as interference, sample sort can interact with the device in an uncontrolled manner. Hence sample sort performs better than external merge sort and MergePass as it avoids repeated data copies to DRAM. However, WiscSort OnePass is still faster due to smaller data movement by dealing with keys, unlike sample sort, which moves records in place.
Due to its reliance solely on sequential reads, merge sort is forced to write the record twice, making it the slowest. Due to the lack of $(I)$ property, we observe that MergePass without Interference-Aware Scheduling (I/O overlap) performs similarly to MergePass with Interference-Aware Scheduling. BRD-Device performance results suggest that the improved random-read bandwidth alone is enough to warrant redesigning the sorting system.

\subsubsection{\textbf{BARD-Device:} Byte Addressable, Asymetric Read-Write, \\Higher Random Read, Device Concurrency} \hfill \\
Newer storage media exhibit eccentric characteristics that require systems to adapt accordingly. One such characteristic PMEM has is the asymmetric read-write performance, where the read bandwidth can be ~$4x$ faster than the write. To mimic a device with such property with a larger asymmetry, we emulate BARD-Device to have writes 500ns slower than reads $(A)$. However, its sequential and random-read bandwidth is the same $(R)$, and it is bye-addressable $(B)$. 

From Figure \ref{fig:cxl-ba-RR-ARW}, we can observe that writes dominate the overall run time. Since sample sort does not suffer from $(I)$, it performs better than WiscSort MergePass; OnePass does slightly better due to its reduced sorting time (sorting only key-pointer). 
As expected, the difference in performance between external merge sort and WiscSort is 2x due to the reduced writes during the run phase.
The MergePass I/O overlap sees similar performance as that of MergePass no I/O overlap, indicating that there will be benefits of Interference-Aware Scheduling only in the presence of read-write interference property. 
As the writes become more costly, the benefits of WiscSort OnePass degrade compared to sample sort, external merge sort, and even WiscSort MergePass. Nevertheless, OnePass still achieves the lowest execution time due to reducing the total amount of data movement.\looseness=-1


\vspace{-1mm}
\section{Discussion and Future Work}

WiscSort converts a row-oriented database to a column-oriented one on the fly, this enables provisions to provide late materialization if required. 
For example, a range of sorted key values can be generated \textit{on demand} with the help of IndexMap files; or two IndexMap files can be used to perform joins on relations without moving entire values associated with them. 
BRAID provide new opportunities for late materialization without requiring complex data structures, warranting us to rethink how HTAP databases must be redesigned from the ground up, just like GDPR-compliant databases \cite{vldb-gdpr, hotstorage-gdpr}.

Awareness of the concurrency constraints plays an important role in maximizing BRAID bandwidth. We assume a single-tenant system, but a single application can have adversarial effects on all other applications' read bandwidth if multiple applications are deployed.
To avoid such degradation, new system-wide I/O schedulers that are both interference and concurrency aware are warranted, i.e., a scheduler that scales requests without degrading bandwidth and provides non-overlapping reads and writes across applications.

WiscSort is currently designed to work with datasets that fit within the BRAID device. However, if the dataset is spilled to secondary storage like SSD/HDD, we need new design approaches that efficiently use all the heterogeneous storage devices simultaneously. Moreover, one could compress the IndexMap files before they are written to the BRAID device to better adhere to the BRAID property. Compression will be worthwhile only if the cost of reads and decompression is smaller than that of compression and writes. Compression also places new demands on the CPU, which must be considered. 
However, these optimizations are orthogonal to our contributions and are left as future work.




\section{Related Work}

A few prior works have looked at algorithms for asymmetric read-write costs on persistent memory \cite{aem-sort, db-rethinking-nvm, write-limit-join, Btree-asymmetric-io}. In general, they overcome the difference by performing multiple reads to reduce the total number of writes made to the device.  For example, the B-tree variant \cite{db-rethinking-nvm} does not sort the keys in a leaf node nor repack a leaf after a deleted key, thereby avoiding the write cost of sorting and repacking, at the expense of additional reads when searching. AEM-sort \cite{aem-sort} introduced a theoretical model to asymptotically reduce the number of writes of multi-way merge sort, sample sort, and heap sort over the original. Finally, various "write-limited" sorting and join algorithms where introduced \cite{write-limit-join} that preferred to scan the input multiple times to reduce the writes. 

MONTRES-NVM \cite{montres-nvm} and NVMSorting \cite{nvm-sorting} introduce techniques to leverage partially sorted inputs on persistent memory. They detect naturally sorted portions of the data set which are ignored during the run generation phase to reduce the total number of writes. These natural runs are merged on the fly during MERGE phase. WiscSort does not make any assumptions about the input data distribution. None of the above algorithms were designed for a real device like PMEM, so they all fail to consider the performance characteristics, such as $(R, I, D)$ of BRAID. Nevertheless, WiscSort is orthogonal to all the above solutions and combining them could further benefit the sorting performance.

\section{Conclusion}
Byte-addressable storage requires application redesign for peak performance.
We introduced the $BRAID$ model to characterize the important properties of byte-addressable storage that applications must comply with. 
We showed that conventional access strategies, such as performing sequential reads and overlapping reads and writes, do not hold well any more.
We proposed WiscSort, a high-performance concurrent sorting system that complies with BRAID using three important features: key-value separation, thread-pool control, and interference-aware scheduling.
Our results show that WiscSort can be 2-7x faster than the competing approaches.


\begin{acks}
We thank the anonymous reviewers for their insightful comments
This material was supported by gifts from Google.
\end{acks}

\clearpage

\bibliographystyle{ACM-Reference-Format}
\bibliography{base}

\end{document}